\documentclass[showpacs,preprintnumbers,amsmath,aps,nofootinbib,amssymb,superscriptaddress,twocolumn]{revtex4}
\usepackage{float}
\usepackage[caption=false,justification=justified]{subfig}  
\usepackage{graphicx}
\usepackage[dvipsnames]{xcolor}
\usepackage[english]{babel}
\usepackage{bm}
\usepackage[font=scriptsize]{caption}
\usepackage{ragged2e}
\usepackage[font=small,labelfont=bf,justification=raggedright]{caption}
\usepackage[utf8]{inputenc}
\usepackage[tikz]{bclogo}
\usepackage[framemethod=tikz]{mdframed}

\newsavebox{\measurebox}
\usepackage[colorlinks=false]{hyperref}
\begin{document}
	\baselineskip 16pt
	\title{Noncommutative effective LQC: A (pre-)inflationary dynamics investigation}
	\author{Luis Rey D\'iaz-Barr\'on}
	\email{lrdiaz@ipn.mx}
	\affiliation{Unidad Profesional Interdisciplinaria de Ingenier\'ia
		Campus Guana\-jua\-to del Instituto Polit\'ecnico Nacional.\\
		Av. Mineral de Valenciana No. 200, Col. Fraccionamiento Industrial Puerto Interior, C.P. 36275, Silao de la Victoria, Guana\-jua\-to,
		M\'exico.}
	
	\author{Abraham Espinoza-Garc\'ia}
	\email{aespinoza@ipn.mx}
	\affiliation{Unidad Profesional Interdisciplinaria de Ingenier\'ia
		Campus Guana\-jua\-to del Instituto Polit\'ecnico Nacional.\\
		Av. Mineral de Valenciana No. 200, Col. Fraccionamiento Industrial Puerto Interior, C.P. 36275, Silao de la Victoria, Guana\-jua\-to,
		M\'exico.}
	
	\author{S. P\'erez-Pay\'an}
	\email{saperezp@ipn.mx}
	\affiliation{Unidad Profesional Interdisciplinaria de Ingenier\'ia
		Campus Guana\-jua\-to del Instituto Polit\'ecnico Nacional.\\
		Av. Mineral de Valenciana No. 200, Col. Fraccionamiento Industrial Puerto Interior, C.P. 36275, Silao de la Victoria, Guana\-jua\-to,
		M\'exico.}
	
	\author{J. Socorro}
	\email{socorro@fisica.ugto.mx}
	\affiliation{Departamento de
		F\'{\i}sica, DCeI, Universidad de Guanajuato-Campus Le\'on, C.P.
		37150, Le\'on, Guanajuato, M\'exico.}
	
	\begin{abstract}
		We conduct a (pre-)inflationary dynamics study within the framework of a simple noncommutative extension of effective loop quantum cosmology---put forward recently by the authors---which preserves its key features (in particular, the quantum bounce is maintained). A thorough investigation shows that the (pre-)inflationary scenario associated to the chaotic quadratic potential is in the overall the same as the one featured in standard loop quantum cosmology (which reinforces the conclusion reached by the authors in a preliminary analysis). Hence, this (pre-)inflationary scenario does not easily distinguish between standard loop quantum cosmology and the aforementioned noncommutative scheme. It is argued that a particular tuning of the noncommutativity parameter could accommodate for subtle effects at the level of primordial perturbations (the hybrid quantization framework being a tentative route of analysis).
	\end{abstract}
	
	
	\maketitle
	
	\section{Introduction}\label{section1}
	Leading approaches to quantum gravity (loop quantum gravity (LQG) and string/M theory being at the moment the most prominent candidates) seem to point to a breakdown of the continuum picture of spacetime, at a fundamental level. In particular, it is anticipated that an accurate description of the very early universe be achieved by appending a certain degree of discreteness to the gravitational field. On one hand, LQG leads to a ``quantum geometry" in which spatial area and volume operators feature a discrete set of eigenvalues whose associated levels ``crowd'' exponentially as the eigenvalues grow (the minimum area eigenvalue being of paramount importance)\cite{reviews-lqg}; on the other, low energy limits of string/M theory lead to noncommutative gauge theories \cite{st-ncg}, \cite{reviews-ncft}. LQG has the appealing of not involving higher dimensions, nor supersymmetry, in its formulation; furthermore, loop quantum cosmology (LQC) (the implementation of the ideas and methods of LQG to cosmologically relevant symmetry-reduced minisuperspace models) has the benefit of leading to a true initial singularity resolution (particular black hole singularities have also been shown to be removed as a consequence of the quantum geometry)\cite{reviews-lqg},\cite{reviews-lqc}. The presumption that the final theory of quantum gravity would display, at an effective level, relevant features of these two leading approaches is not too adventurous, given their claimed relative successes. A rather simple and direct avenue to explore such blending in a single effective framework is the implementation of a simplified form of noncommutativity at the effective scheme of LQC (a recent review on implementation of noncommutativity-related ideas in cosmology is given in \cite{ncqc-review}). Of course, it is most desirable that this construction retains some of the prominent results of LQC (most notably, the initial singularity resolution and compatibility with the inflation mechanism). Indeed, the authors of the present investigation have shown, in a preliminary analysis, that when considering a momentum space $\theta-$deformation at the effective scheme of LQC (with a quadratic potential scalar field) there exist solutions that are singularity free and at the same time display a sufficiently long inflationary period \cite{prd}. Such preliminary results make plausible carrying a more careful study of the (pre-)inflationary dynamics within this noncommutative extension of effective LQC. The main purpose of this letter is the undertaking of this more thorough study. 
	
	In the following, we review the basics of both, the effective scheme of LQC and the noncommutativity paradigm as implemented within the minisuperspace approximation. 
	\subsection{Basics of effective LQC}
	We begin by spelling out some facts pertaining the effective scheme of LQC (detailed and rather thorough accounts can be found in the standard reviews \cite{reviews-lqc}). As already stated, LQC is the quantization of symmetry-reduced general relativity, following closely the ideas and methods of LQG. Succinctly, the usual 3-metric Hamiltonian formulation of general relativity is rewritten (by means of a canonical transformation) in terms of the SU(2) Ashtekar-Barbero connection, therefore casting General Relativity (GR) as a gauge theory. Further, careful implementation of canonical quantization requires the holonomies of the Ashtekar-Barbero connection to be considered as the true configuration field variable (instead of the connection itself). Accordingly, holonomies of a highly symmetric sector (e.g. the homogeneous sector) of the connection superspace are regarded as the cosmologically relevant configuration field variable. Canonical quantization is then implemented on such connection minisuperspace, the resulting quantum framework is known as loop quantum cosmology. Hence, LQC suffers from the same minisuperspace approximation drawback as standard quantum cosmology, i.e. LQC is not a symmetry-reduced LQG. In this respect, it is important to note that, in particular, the noncommutativity among the conjugate momenta fields (electric fluxes) in full LQG is completely lost in LQC, signaling possible significant loss of information when carrying the minisuperspace approximation.
	
	One of the main results of LQC is the avoidance of the initial singularity via a quantum bounce. The quantum corrections can be seen as holonomy- and inverse triad-related, with the former being more relevant. The standard effective scheme of LQC is achieved by focusing on such holonomy-related quantum corrections. In the isotropic case, the classical geometro-dynamical variables which are selected by (effective) LQC are $\beta=\gamma\dot{a}/(aN)$ and $V=a^3$, satisfying 
	\begin{equation}
		\{\beta,V\}=4\pi G\gamma,\quad 
		\mathsf{P}(y)=4\pi G\gamma\, \partial_{\beta}\wedge\partial_{V}\label{PB},
	\end{equation}
	where $\gamma$ is the so called Barbaro-Immirzi parameter and $\mathsf P(y)$ denotes the Poisson structure in variables $y=(\beta, V)$. The (gravitational) Hamiltonian is written in these variables as ($N=1$)
	\begin{equation}
		\mathcal{H}=-\frac{3}{8\pi G\gamma^{2}}\beta^2V.
		\label{ham1}
	\end{equation}
	It has been shown (via analytical \cite{reviews-lqc} and numerical \cite{review-numlqc} investigations) that (in the isotropic case) the quantum corrections associated to LQC are very well approximated by performing the \textit{replacement} 
	\begin{equation}
		\beta\to\frac{\sin{\lambda\beta}}{\lambda},
	\end{equation}
	within the classical Hamiltonian, where $\lambda^{2}=4\sqrt{3}\pi\gamma\ell^{2}_{p}$ is the lowest eigenvalue (corresponding to quantum states compatible with the assumed spatially homogeneous and isotropic geometry) of the area operator in the full loop quantum gravity---and where $\ell_p$ is the Planck lenght. The resulting effective Hamiltonian is given by
	\begin{equation}
		\mathcal{H}_{\mathrm{eff}}=-\frac{3}{8\pi G\gamma^{2}\lambda^{2}}\sin^{2}(\lambda\beta)V+\frac{p^{2}_{\phi}}{2V}.
		\label{ham2}
	\end{equation}
	Thus, the effective scheme of LQC supports meaningful quantum corrections within a purely classical setup. This situation has  allowed to carry exploratory investigations of quantum effects on a wide range of cosmological backgrounds in this rather simplified theoretical framework.   
	
	The equations of motion associated to the ``holonomized'' Hamiltonian \eqref{ham2} with the standard Poisson structure 
	\begin{equation}
		\mathsf{P}(y)=4\pi G\gamma\, \partial_{\beta}\wedge\partial_{V}+\partial_{\phi}\wedge\partial_{p_\phi},\quad y=(\beta,V,\phi,p_\phi),
		\label{PB2}
	\end{equation}
	are
	\begin{align}
		&\dot{\beta}=4\pi G\gamma\frac{\partial\mathcal{H}_{\mathrm{eff}}}{\partial V}=-\frac{3}{\gamma\lambda^{2}}\sin^{2}(\lambda\beta),\label{eom-beta}\\
		&\dot{V}=-4\pi G\gamma\frac{\partial\mathcal{H}_{\mathrm{eff}}}{\partial\beta}=\frac{3}{\gamma\lambda}V\sin(\lambda\beta)\cos(\lambda\beta),\label{eom-vol}\\
		&\dot{\phi}=\frac{\partial\mathcal{H}_{\mathrm{eff}}}{\partial p_{\phi}}=\frac{p_{\phi}}{V},\label{eom-phi}\\
		&\dot{p}_{\phi}=-\frac{\partial\mathcal{H}_{\mathrm{eff}}}{\partial\phi}=0.\label{eom-p-phi}
	\end{align}
	Since $V=a^{3}$, $\frac{\dot{V}}{3V}=\frac{\dot{a}}{a}=:H$, then, taking into account the effective Hamiltonian constraint, $\mathcal{H}_{\mathrm{eff}}=0$, and the equation of motion for $V$, we have,
	\begin{equation}
		H^{2} :=\left(\frac{\dot{V}}{3V}\right)^{2}=\frac{8\pi G}{3}\rho\left(1-\frac{\rho}{\rho_{\mathrm{c}}}\right),
		\label{mod-friedmann}
	\end{equation}
	where 
	\begin{equation}
		\rho=\frac{\dot\phi^2}{2}=\frac{p^{2}_{\phi}}{2V^{2}}=\frac{3}{8\pi G\gamma^{2}\lambda^{2}}\sin^{2}(\lambda\beta),
	\end{equation} 
	and $\rho_{{c}}$ is the maximum value that $\rho$ can take in view of the effective Hamiltonian constraint, 
	\begin{equation}
		\rho_{{c}}=\frac{3}{8\pi G\gamma^{2}\lambda^{2}}\simeq0.41\rho_{\mathrm p},
	\end{equation}
	with $\rho_{\mathrm p}$ the Planck density. Equation \eqref{mod-friedmann} is the so called \textit{modified Friedmann equation}.
	
	In full LQC it is shown that $\beta$ takes values on the interval $\left(0,\frac{\pi}{\lambda}\right)$ \cite{lqc-status-report}, therefore, the volume function reaches a stationary point at $\beta=\frac{\pi}{2\lambda}$. From the equation of motion (\ref{eom-beta}) it is established that $\beta$ is a decreasing function of time ($\dot{\beta}\leq0$), which in turn implies that such stationary point is only reached once (and that the maximum of the energy density function is also attained only once). From this monotonic behavior of $\beta$ it also follows that such stationary point corresponds to a minimum (since $\rho$ is always bounded, the stationary point associated to a vanishing volume does not feature in the solutions ofx the equations of motion). Thus, in the effective scheme of LQC, a minimum of the volume function is reached precisely once, and this minimum volume corresponds to the maximum of the energy density function. This state of affairs is summarized by the statement that singularity resolution is achieved through a (minimum-volume) quantum bounce. As already stated, this singularity resolution is one of the landmark features of LQC, and in this effective scheme it is generic: all solutions undergo a minimum-volume bounce exactly once. One could argue that singularity resolution is achieved due to the semiclassical assumptions underlying this effective scheme, but it has been shown that the quantum bounce is likewise ``robust'' in the full LQC (of the flat FLRW model with a free scalar field) \cite{robustness}. 
	
	Note that the modified Friedmann equation (\ref{mod-friedmann}) incorporates in a rather simple way the occurrence of the bounce (at $\rho=\rho_{\mathrm{c}}$). In the limit $\lambda\rightarrow0$ (no area gap) we recover the ordinary Friedmann equation ${H}^{2}=\frac{8\pi G}{3}\rho$.
	
	By making use of the Hamiltonian constraint, the equation for $\beta$ can be decoupled and solved by direct integration. Upon substitution of the solution $\beta=\beta(t)$ in (\ref{eom-vol}) we find the corresponding solution for $V=V(t)$. Similarly, we substitute the solution for $V=V(t)$ in (\ref{eom-phi}) to get the solution $\phi=\phi(t)$. We have,
	\begin{align}
		&\beta(t)=\frac{1}{\lambda}\mathrm{arccot}\left(\frac{3t}{\gamma\lambda}\right),\label{sol-beta-lqc}\\
		&V(t)=C_1\sqrt{\gamma^{2}\lambda^{2}+9t^{2}},\label{sol-v-lqc}\\
		&\phi(t)=C_2+\frac{p_{\phi}}{3C_1}\mathrm{log}\left(3t+\sqrt{\gamma^2\lambda^2+9t^2}\right),\label{sol-fi-lqc}\\
		&p_\phi(t)=p_\phi.
	\end{align}
	From (\ref{sol-beta-lqc}) we note that the solution is consistent with $\beta\in\left(0,\frac{\pi}{\lambda}\right)$ (by considering another ``branch'' we get $\beta\in\left[\left.-\frac{\pi}{2\lambda},\frac{\pi}{2\lambda}\right.\right)$, which is also employed in the literature of effective LQC (see, for instance, \cite{corichi-tatiana}).
	\subsection{Basics of the noncommutativity paradigm}
	We now turn to state the main ideas of the noncommutativity paradigm. Let us first recall some basic facts of the geometrical formulation of classical mechanics.
	Remember that the phase space $\Gamma=T^\star\mathcal{Q}$ (the cotangent bundle of $\mathcal{Q}$), where $\mathcal{Q}$ is the configuration space of a mechanical system, is a symplectic manifold, and that the inverse $\mathsf P$ of the symplectic structure $\omega$ ($\omega_{\alpha\beta}\mathsf P^{\beta\gamma}=\delta^{\gamma}_{\alpha}$) defines the Poisson bracket $\{\cdot,\cdot\}$ (which endows $\Gamma$ with a Lie algebra structure). The equations of motion   
	\begin{equation}
		\dot{q}^a=\partial_{p_a}\mathcal{H},\quad\dot{p}_a=-\partial_{q^a}\mathcal{H} \label{local-ham-sys}
	\end{equation}
	of the corresponding mechanical system $(\Gamma,\mathcal{H})$ ($\mathcal{H}$ being the Hamiltonian of the system) can be written in a purely geometric coordinate-independent way,
	\begin{equation}
		\dot{x}^\alpha=\mathsf{P}^{\alpha\beta}(d\mathcal{H})_\beta.
	\end{equation}
	
	In canonical coordinates $x=(x^1,...,x^{2n})=(q^1,q^2,...,q^n,p_1,p_2,...,p_n)$ (i.e. in a Darboux chart), the symplectic structure acquires the usual form ($a=1,...,n$)
	\begin{equation}
		\omega=J_{\alpha\beta}dx^{\alpha}\wedge dx^{\beta}=dx^{n+a}\wedge dx^{a}=dp_{a}\wedge dq^{a}.
	\end{equation}
		We therefore have (in canonical coordinates)
		\begin{equation}
			\mathsf{P}=J^{\alpha\beta}\partial_{\alpha}\wedge\partial_{\beta}=\partial_a\wedge\partial_{n+a}=\partial_{q^{a}}\wedge\partial_{p_{a}}.\label{flat-structure}
		\end{equation}
			(Notation $J_{\alpha\beta}$, $J^{\alpha\beta}$ is usually reserved for the components of the symplectic and Poisson structures when expressed in a Darboux chart.)
			
			Poisson brackets are defined in terms of the Poisson structure by 
			\begin{equation}
				\{f,g\}:=\mathsf{P}(df,dg)=\mathsf{P}^{\alpha\beta}df_\alpha dg_\beta.
			\end{equation}
			In particular, for the canonical coordinates $x=(x^1,...,x^{2n})=(q^1,q^2,...,q^n,p_1,p_2,...,p_n)$, we have the standard relations
			\begin{equation}
				\{q^{a},p_{b}\}=\delta^{a}_{b},\qquad\{q^{a},q^{b}\}=\{p_{a},p_{b}\}=0.
				\label{c-flat relations}
			\end{equation}
			
			The noncommutativity paradigm we consider is closely related to the so called quantum mechanics on phase space (QMPS), which provides an alternate way to consistently formulate quantum mechanics (the recent monograph \cite{curtright} provides a comprehensive account on QMPS). 
			
			Roughly, in QMPS the canonical commutation relations 
			\begin{equation}
				[\hat{q}^i,\hat{p}_j]=i\hbar\delta^i_j,\quad[\hat{q}^i,\hat{q}^j]=[\hat{p}_i,\hat{p}_j]=0\label{cancommrel}
			\end{equation}
			are realized by means of a \textit{deformation} of the usual point-wise multiplication of phase space functions, $f\cdot g\mapsto f\star g$. This \textit{star-product} ($\star$-product) can be thought of as a ``power series'' of $\frac{i\hbar}{2}\mathsf{P}(df,dg)$ (where $\mathsf{P}$ is the Poisson structure of classical phase space) in which the ``zeroth power'' is the usual commutative product $f\cdot g$. We can therefore obtain standard classical mechanics by simply taking $\hbar\to0$. By considering the (classical) commutator $[f,g]_\star=f\star g-g\star f$, the usual canonical commutation relations \eqref{cancommrel} are realized in \textit{classical} phase space $\Gamma$. Central to the scheme of QMPS is the so called \textit{Wigner function}, which is a quasi-probability distribution in phase space whose purpose was to provide quantum corrections to the classical Liouville measure of classical statistical mechanics. In QMPS probability amplitudes are calculated employing the Wigner function.
			
			The connection with canonical quantum mechanics is given by working in the well known Weyl representation (unitarily equivalent to the standard Schr\"odinger representation) in which the canonical commutation relations \eqref{cancommrel} acquire an ``exponentiated form''; this representation is encoded in a quantization prescription, the so called Weyl map $\mathcal W$, which associates a (symmetric ordered) quantum operator $\hat{f}$ to a phase space function $f$. The Weyl map $\mathcal W$ is invertible ($\mathcal{W}^{-1}(\hat{f})$ is called the Weyl symbol of $\hat{f}$ --it is also called the Wigner transform of $\hat{f}$--). The Wigner function maps (up to multiplicative factors of $\hbar$) to the density operator $\hat{\rho}$ under the Weyl transform (i.e. the Wigner function is the Wigner transform of the density operator). The $\star$-product is the image of the operator product (in the Weyl representation) under the Wigner transform, $f\star g=\mathcal{W}^{-1}(\mathcal{W}(f)\mathcal{W}(g))$.
			
			More explicitly, the Weyl-Wigner-Moyal (WWM) correspondence relates a phase space function $f$ with its operator counterpart $\mathcal{W}(f)=\hat{f}$, via the following map 
			\begin{align}
				&f\mapsto\mathcal{W}(f)=\hat{f}(\hat{q},\hat{p}),\quad\mathrm{with}\nonumber\\
				&\hat{f}(\hat{q},\hat{p})=\int \tilde{f}(\xi,\eta)\exp\left({\frac{i}{\hbar}\left(\hat{p}_{a}\xi^{a}+\hat{q}^{a}\eta_{a}\right)}\right)w(\xi,\eta)d^{l}\xi d^{l}\eta,
				\label{wwm}
			\end{align}
			where $\tilde{f}$ is the Fourier transform of $f$, $w$ a weight function, and $\dim(T^{\star}\mathcal{Q})=2l$; with $\hat{q}^{a},\hat{p}_{b}$ satisfying the canonical Heisenberg algebra \eqref{cancommrel}.
			We define the $\star$-product with the help of this correspondence by
			\begin{equation}
				\mathcal{W}(f\star g)=\mathcal{W}(f)\mathcal{W}(g)
				\label{star-product}
			\end{equation}
			that is, $f\star g$ is the Weyl symbol of $\hat{f}\hat{g}$. For this case of the Heisenberg algebra (\ref{cancommrel}) and in the symmetric ordering ($w=1$) we have
			\begin{align}
				f\star g&=\exp\left(\frac{i\hbar}{2}\mathsf P(df,dg)\right)\nonumber\\
				&=fg+\sum_{r=1}^{\infty}\left(\frac{i\hbar}{2}\right)^{r}\frac{1}{r!}\mathsf P^{r}(df,dg)\nonumber\\
				&=fg+\frac{i\hbar}{2}\{f,g\}+\mathcal{O}(\hbar^{2})
				\label{moyal}
			\end{align}
			where in the expansion, terms of the form $\mathsf{P}^{r}(df,dg)=J^{\mu_{1}\nu_{1}}\cdot\cdot\cdot J^{\mu_{r}\nu_{r}}\left(\partial_{\mu_{1}...\mu_{r}}f\right)\left(\partial_{\nu_{1}...\nu_{r}}g\right)$ (the ``rth power'' of the Poisson bracket of $f$ and $g$) are to be considered. This is the Moyal $\star$-product \cite{moyal}, it replaces the ordinary point-wise multiplication in the algebra of functions defined in phase space. The Moyal $\star$-bracket $[f,g]_\star:=f\star g-g\star f$ is thus responsible for the realization of the canonical Heisenberg algebra (\ref{cancommrel}). 
			This product, together with the Wigner function, is the cornerstone of QMPS.
			
			The noncommutativity paradigm employs some of the ideas and methods of QMPS (most notably, the $\star$-product), and can therefore be viewed from the perspective of the QMPS scheme: Given more general commutation relations, e.g. \eqref{theta-def},
			construct the corresponding $\star$-product by accordingly considering an appropriate modification $\tilde{\mathsf{P}}$ to $\mathsf{P}$. This $\tilde{\mathsf{P}}$ will be such that the classical Poisson relations corresponding to the above commutation relations are fulfilled (in the same dynamical variables). Once the $\star$-product is constructed, the advancement of a noncommutative quantum mechanics in phase space (NCQMPS) can be pursued. Instead, we view \eqref{theta-def} as an extension to standard quantum mechanics, resulting in a more noncommutative framework than the original standard one. The implementation of such additional noncommutativity is achieved by implementing a $\star$-product on \textit{quantum} phase space. This point of view results in a framework known as noncommutative quantum mechanics (NCQM) \cite{mezin}.
			
			The introduction of the modification $\tilde{\mathsf{P}}$ can, at the classical level, be interpreted as a manifestation of the phase space coordinates not being canonical, which in turn would imply that a new phase space $\tilde\Gamma$ is being considered (a ``deformation'' to the original one, $\Gamma$). An interesting point of view is the assumption that this new $\tilde\Gamma$ \textit{is the true phase space of the mechanical system} under study, and that it was initially overlooked due to the noncommutativity scale being unnoticed, that is, the mechanical system $(\tilde\Gamma,\mathcal{H})$ is considered as the true mechanical system, instead of $(\Gamma,\mathcal{H})$. This phase space $\tilde\Gamma$ would therefore incorporate (noncommutative) quantum corrections in a purely classical setup. In this interpretation, the classical equations of motion would not be in canonical form in the original (now non-canonical) variables $(y)$ (the Hamiltonian function $\mathcal{H}$ being the same). This deviation from the canonical form of the equations of motion (\textit{noncommutative equations of motion}) would therefore encode such quantum corrections. Due to the Darboux theorem, one would be able to find a new chart with local coordinates $x$ (via a transformation $y=\mathsf T(x)$) in which $\tilde{\mathsf{P}}$ acquires the canonical form \eqref{flat-structure} (so that the classical equations of motion would be in canonical standard form), but the Hamiltonian function would be expressed as $\mathcal{H}(y)\mapsto\mathcal{H}(x)=\mathcal{H}(\mathsf{T}(x))$; the quantum corrections from noncommutativity would in this case be encoded in the Hamiltonian $\mathcal{H}(x)$. In our investigations, when performing computations, we prefer to work in canonical coordinates $(x)$ than with the non canonical coordinates $(y)$ (which were canonical in the original mechanical system $(\Gamma,\mathcal{H})$).
			
			Consider for instance the deformed commutation relations (``$\theta-$deformation'' on momentum sector)
			\begin{equation}
				[\hat{q}^{a},\hat{p}_{b}]=i\hbar\delta^{a}_{b},\quad[\hat{p}_{a},\hat{p}_{b}]=i\theta_{ab},
				\label{theta-def}
			\end{equation}
			with $\theta_{ab}$ an antisymmetric constant real matrix. 
			
			In light of the discussion above, quantum corrections associated to \eqref{theta-def} could be incorporated in a classical setup by considering the associated ``deformed'' mechanical system $(\tilde\Gamma,\mathcal{H})$, for which the Poisson structure would take the form 
			\begin{equation}
				\mathsf{P}(y)=\partial_{y^a}\wedge\partial_{y^{n+a}}+\theta_{ab}\,\partial_{y^{n+a}}\wedge\partial_{y^{n+b}}~~{\rm with}~~(a<b)
				\label{q-def-ncqm}
			\end{equation}
			where $y=(q^1,q^2,...,q^n,p_1,p_2,...,p_n)$ are the phase space coordinates used for the description of the \textit{original} mechanical system $(\Gamma, \mathcal{H})$. This concludes our succinct account on the basics of LQC and $\theta-$deformed (quantum) mechanics.

			The remaining of the manuscript is displayed as follows. Section (\ref{section2})) is devoted to quickly reviewing the model presented in \cite{prd}. A careful analysis of the corresponding post-bounce and pre-inflationary dynamics is presented in section (\ref{section3}). In section (\ref{section4})) we discuss the main results and spell out some perspectives and undergoing work. 
			
			\section{A Noncommutative effective lqc}\label{section2}
		The contents of this section are based on certain specific parts of Refs. \cite{hindawi}, \cite{ijmpd} and \cite{prd}. Its purpose is to make the document somewhat self-contained, and to highlight certain aspects which will be of relevance in the development of the main part of the manuscript (Section \ref{section3}).
		\subsection{The free case}
		
		Consider the mechanical system $(\Gamma, \mathcal{H})$ defined by \eqref{PB2} and \eqref{ham2}. We would like to consider a new mechanical system $(\tilde\Gamma, \mathcal{H})$ which incorporates quantum corrections from an underlying $\theta-$deformation analogous to \eqref{theta-def}. The Poisson algebra of this new $(\tilde\Gamma,\mathcal{H})$ is therefore given by
		\begin{equation}
			\{p^{{nc}}_{\phi},V^{{nc}}\}=\theta,\quad 
			\{\beta^{{nc}},V^{{nc}}\}=4\pi G\gamma,\quad 
			\{\phi^{{nc}},p^{{nc}}_{\phi}\}=1,
			\label{nc-mom}
		\end{equation}
		with the remaining brackets zero as usual. The corresponding Poisson structure $\tilde{\mathsf{P}}$ is therefore expressed as  
		\begin{equation}
			\tilde{\mathsf{P}}(y)=4\pi G\gamma\,\partial_\beta\wedge\partial_V+\partial_\phi\wedge\partial_{p_\phi}+\theta\,\partial_{p_\phi}\wedge\partial_V.
			\label{nc-mom2}
		\end{equation}
		Here, variables $y=(\beta^{{nc}},V^{{nc}},\phi^{{nc}},p_\phi^{{nc}})$ are the same variables which were canonical in the original system $(\Gamma, \mathcal{H})$ defined by \eqref{ham2} and \eqref{PB2} (we append the label ``${nc}$'' in order to remind us that they are non canonical variables in the new system $(\tilde\Gamma,\mathcal{H})$).
		
		Now, the relations
		\begin{equation}
			V^{{nc}}=V+a\theta\phi,\quad p_{\phi}^{{nc}}=p_\phi+b\theta\beta,\quad \beta^{{nc}}=\beta,\quad\phi^{\mathrm{nc}}=\phi,
			\label{nc-rel-mom}
		\end{equation}
		with $4\pi G\gamma b-a=1$, define a family of phase space coordinate transformations $x=\mathsf{T}(y)$ to \textit{canonical} variables $x=(\beta,\phi,V,p_\phi)$ for system $(\tilde\Gamma,\mathcal{H})$. As already stated, we prefer to work with canonical coordinates so that the equations of motion take the standard form. Accordingly, the equations of motion associated to the Hamiltonian,
		\begin{equation}
			\mathcal{H}^{{nc}}_{\mathrm{eff}}=-\frac{3}{8\pi G\gamma^{2}\lambda^{2}}\sin^{2}(\lambda\beta)(V+a\theta\phi)+\frac{\left(p_\phi+b\theta\beta\right)^{2}}{2(V+a\theta\phi)},
			\label{nc2-frw-ham}
		\end{equation}
		are to be computed in the standard way for variables $x=(\beta,\phi,V,p_\phi)$. Such equations are,
		\begin{align}
			\dot{\beta}=&4\pi G\gamma\frac{\partial\mathcal{H}^{{nc}}_{\mathrm{eff}}}{\partial V}=-\frac{3}{\gamma\lambda^{2}}\sin^{2}(\lambda\beta),\\
			\dot{V}=&-4\pi G\gamma\frac{\partial\mathcal{H}^{{nc}}_{\mathrm{eff}}}{\partial\beta}=\frac{3}{\gamma\lambda}(V+a\theta\phi)\sin(\lambda\beta)\cos(\lambda\beta)\nonumber\\
			&-4\pi G\gamma b\theta \frac{p_\phi+b\theta\beta}{V+a\theta\phi},\\
			\dot{\phi}=&\frac{\partial\mathcal{H}^{{nc}}_{\mathrm{eff}}}{\partial p_{\phi}}=\frac{p_\phi+b\theta\beta}{V+a\theta\phi},\\
			\dot{p}_{\phi}=&-\frac{\partial\mathcal{H}^{{nc}}_{\mathrm{eff}}}{\partial\phi}=-\frac{3a\theta}{8\pi G\gamma^{2}\lambda^{2}}\sin^{2}(\lambda\beta)\nonumber\\
			&-a\theta\frac{\left(p_\phi+b\theta\beta\right)^{2}}{2\left(V+a\theta\phi\right)^{2}}.
		\end{align}
		We recover the usual (commutative) equations of motion by taking $\theta\rightarrow0$.
		
		We note that the equation for $\beta$ is the same as in the commutative case, so that $\beta(t)=\frac{1}{\lambda}\mathrm{arccot}\left(\frac{3t}{\gamma\lambda}\right)$.  We also see that it is the case that $\beta$ is still a decreasing function of time (taking values on $\left(0,\frac{\pi}{\lambda}\right)$).
		
		From the Hamiltonian constraint and the equation of motion for $\phi$ it follows that the energy density $\rho$ is exactly the same as in the standard case: 
		\begin{equation}
			\rho=\frac{\dot\phi^2}{2}=\rho_{\mathrm c}\sin^2{\lambda\beta},
		\end{equation}
		and so it is still bounded and it attains the same maximum value $\rho_c$ as in the commutative case. We emphasize that the energy density function not only has the same functional form in terms of $\beta$ as in the usual case, but that it is exactly the same function of time, since the solution $\beta=\beta(t)$ is the same as in the standard commutative case. 
		
		The equation of motion for $V$ indicates that several stationary points of the volume function are \textit{in principle} possible, and therefore the volume function could attain minima not necessarily all positive, and not necessarily associated to the maximum of the energy density function. This scenario would be in striking contrast with the standard commutative case, where all volume solutions undergo precisely one minimum, and this minimum is always related to the maximum in the energy density. 
		
		In order to be able to simplify the analysis required to establish wether or not a single minimum-volume bounce is achieved, we restrict to the canonical coordinates $x=(\beta,\phi,V,p_\phi)$ given by the particular chart defined by relations \eqref{nc-rel-mom} with $b=0$ ($a=-1$). It is important to stress that this choice merely amounts to fixing a Darboux chart in which further study of the mechanical system will be carried. Working in this particular chart we observe that by requiring $V^{{nc}}(t_c)>0$ a single minimum-volume bounce (associated to the maximum of the energy density) is guaranteed to take place.
		Indeed, if such condition is fulfilled, we have $\dot{V}(t_c)=0,\ \ddot{V}(t_c)>0$ at $\rho(t_c)=\rho_c$. Furthermore, given the monotonic behavior of $\beta$ and the bounded character of $\rho$, this would be the only stationary point of $V$. In other words, by studying the system in an appropriate chart, we have shown that the occurrence of a (single) big bounce at $t=t_c$ is assured for all solutions.
		
		The solutions for the remaining functions are,
		\begin{align}
			V(t)=&A\sqrt{9t^2+\gamma^2\lambda^2}-\frac{\theta(4\pi\gamma bG-a)}{\lambda\sqrt{12\pi G}}\arctan\left(\frac{3t}{\gamma\lambda}\right)\nonumber\\
			&+\frac{a\gamma\theta}{\sqrt{12\pi G}}\frac{\log(\sqrt{9t^2+\gamma^2\lambda^2}+3t)}{\sqrt{9t^2+\gamma^2\lambda^2}},\\
			\phi(t)=&\frac{\gamma}{\sqrt{12\pi G}}\log{(\sqrt{9t^2+\gamma^2\lambda^2}+3t)}.
		\end{align}
		
		\subsection{Inclusion of a generic potential term}
		Consider again the algebra \eqref{nc-mom}, realized by working in canonical variables defined by the family of transformations \eqref{nc-rel-mom}, and a scalar field $\phi$ with generic potential term $\mathcal{V}(\phi)$. 
		
		The deformed Hamiltonian takes the form 
		\begin{equation}
			\mathcal{H}^{{nc}}_{\mathrm{eff}}=-\frac{3}{8\pi G\gamma^3\lambda^2}\sin^2(\lambda\beta)V^{{nc}}+\frac{(p_\phi^{nc})^2}{2V^{{nc}}}+\mathcal{V(\phi)}V^{{nc}}.\label{ncham}
		\end{equation}
		The respective noncommutative equations of motion are (note that both, the Hamiltonian and the equations are expressions involving the canonical variables $x=(\beta,\phi,V,p_\phi)$, we only substitute the non-canonical variables on the right hand side as a mean to abbreviate the expressions),
		\begin{eqnarray}
			\label{feg1} \dot{\beta}=4\pi G\gamma\left[-\frac{3}{4\pi G\gamma^2\lambda^2}\sin^2(\lambda\beta)+2\mathcal{V(\phi)}\right],\\
			\label{feg2} \dot{V}=\frac{3}{\gamma\lambda}V^{{nc}}\sin(\lambda\beta)\cos(\lambda\beta)-\frac{4\pi G\gamma b\theta p_{\phi}^{{nc}}}{V^{{nc}}},\\
			\label{feg3} \dot{\phi}=\frac{p_{\phi}^{{nc}}}{V^{{nc}}},\\
			\label{feg4} \dot{p}_{\phi}=\frac{3a\theta}{4\pi G\gamma^{2}\lambda^{2}}\sin^{2}(\lambda\beta)-2a\theta \mathcal{V(\phi)}-V^{{nc}}\partial_\phi \mathcal{V(\phi)}.
		\end{eqnarray}
		We observe that in the limit $\theta\rightarrow0$ the commutative field equations are recovered. From the equation of motion for $\phi$ and the constraint $\mathcal{H}_{\mathrm{eff}}^{{nc}}=0$, we can see that the energy density takes the form $\rho=\dot\phi^2/2+V(\phi)=\rho_c\sin^2(\lambda\beta)$, which is the same functional form (in terms of $\beta$) as in the standard case, and therefore attains the same maximum value $\rho_c$. However, the functional form in terms of $t$ would not be in general the same as the standard case, since the solution of $\beta(t)$ is likely to depend on the noncommutative parameter $\theta$ (nonetheless, given the monotonic behavior of $\beta$, it is expected that the behavior of the energy density function is in the overall the same as in the standard case). It is also noted that several stationary values of the volume function could in principle be attained, which in turn could result in several bouncing scenarios (depending on the evolution of all dynamical variables).
		
		Guided by the free case, we consider again
		\begin{equation}
			b=0,\ \ V^{{nc}}(t_c)>0 \label{condb=0}
		\end{equation}
		(where, as before, $t_c$ is such that $\rho(t_c)=\rho_c$). The fulfillment of \eqref{condb=0} is a sufficient condition for a minimum volume bounce to be reached when the maximum $\rho_{\mathrm c}$ of the energy density $\rho$ is attained. That is to say, condition (\ref{condb=0}) implies that the standard LQC big bounce is preserved. This will be the only bouncing scenario since $\beta$ is a decreasing function of time (as  (\ref{feg1}) indicates), taking values on $\left(0,\frac{\pi}{\lambda}\right)$, and since $V^{{nc}}=0$ is not allowed during evolution (in view of the equation of motion for $\phi$ and the energy density being bounded).
		
		On the other hand, condition
		\begin{equation}
			\dot{\phi}(t_c)=0,\  b\theta\ddot\phi(t_c)<0 \label{condbounce}
		\end{equation}
		which translates into
		\begin{equation}
			\rho(\phi(t_c))=\mathcal{V}(\phi(t_c)),\  b\theta\ddot\phi(t_c)<0,
		\end{equation}
		is also a sufficient condition for the occurrence of a minimum volume bounce. However, it does not precludes the existence of additional bounces (which could lead to additional minima of volume, not necessarily all positive), hence, singularity resolution is not guaranteed to always be achieved in this latter scenario. 
			Condition (\ref{condbounce}) would allow to fix the value of $\phi$ at this minimum volume bounce. Furthermore, from the equation of motion for $\phi$ it is observed that $p_\phi(t_c)=-b\theta\beta(t_c)=-\frac{b\theta\pi}{2\lambda}$ must hold. We are facing the following situation: by demanding (through condition \eqref{condbounce}) that a minimum volume bounce takes place when the energy density reaches its maximum value, initial conditions for all dynamical variables (except for the volume) get fixed at the bounce. This condition therefore leads to a dramatically reduced number of  solutions which feature a bounce resembling the (single) big bounce of standard LQC. This state of affairs is obviously not particularly appealing.

			For the scalar field we can construct the modified Klein-Gordon equation using (\ref{feg1})-(\ref{feg4}):
			\begin{equation}
				\label{mkge}
				\ddot\phi+\frac{1}{V^{{nc}}}\left[\dot V^{{nc}}+a\theta\phi\right]\dot\phi+\partial_\phi\mathcal{V}(\phi)=0,
			\end{equation}
			we can see that in the limit $\theta\to0$ we recover the commutative Klein-Gordon equation for the scalar field.  
			
			As already stated, in \cite{prd} we have also shown that this noncommutative extension to effective LQC supports universes which (apart from avoiding the initial singularity by means of a single minimum-volume bounce) exhibit a sufficiently long period of inflation for $\mathcal V(\phi)=1/2m\phi^2$.
			\section{Inflation in Noncommutative Effective LQC}\label{section3}
			
			In this section we will focus on post-bounce and pre-inflationary dynamics by specifying a potential $\mathcal{V}(\phi)$. Specifically, we consider the chaotic potential $\mathcal{V}(\phi)=\frac{1}{2}m^2\phi$, where $m= 1.26\times10^{-6}~m_{pl}$. With the specified potential we solve the field equations (\ref{feg1}-\ref{feg4}) in the same way as in \cite{prd}, for a given set of initial conditions $\phi_B=\phi(t_B)$, $\dot\phi_B=\dot\phi(t_B)$, where $t_B$ is the time at the bounce. In the initial investigation \cite{prd} a $\theta-$deformation in the momentum sector of phase-space was shown to support an inflationary period. In this section we will analyze such inflationary model in more detail.
			
			Before beginning with the analysis, we will introduce some essential ingredients related to inflation. 
			
			\begin{itemize}
				\item We will denote the ratio of the potential energy to the critical density by 
				\begin{equation}
					\label{Fb}
					F_B=\frac{\mathcal{V}(\phi_B)}{\rho_c},
				\end{equation}
				depending on the initial conditions, $F_B$ takes values between $[0,1]$, and as in \cite{sloan}, depending on the values that $F_B$ takes, we can define different energy stages after the bounce. If $F_B<10^{-4}$, we say that we are in the extreme kinetic energy domination (EKED) regime; $10^{-4}<F_B<0.5$ characterizes the kinetic energy domination (KED) regime; the potential energy domination (PED) regime is attained by $0.5<F_B<1$.
				
				\item The equation of state parameter $\omega_\phi$ of the scalar field is given by
				\begin{equation}
					\label{omegastate}
					\omega_\phi=\frac{P(t)}{\rho(t)}=\frac{\frac{1}{2}\dot\phi-\mathcal{V}(\phi)}{\frac{1}{2}\dot\phi+\mathcal{V}(\phi)},
				\end{equation}
				it can take values in the interval $[-1,1]$. The values corresponding to $\omega_\phi=-1$ give a PED regime, while for $\omega_\phi=1$ a KED is achieved. The equation of state parameter allows us to divide the evolution of the scalar field before reheating into three stages, similar to LQC cases \cite{Parampreet,piulqc}, these stages are: the bouncing stage, transitions stage and slow roll inflation stage.
				
				\item In the bouncing stage dynamics there is a period of super inflation (SI), which corresponds to an increasing behavior for the Hubble parameter. This epoch begins at the bounce and ends when the Hubble parameter reaches its maximum value ($\dot H=0$). 
				
				\item The slow-roll Hubble parameters $\epsilon_H$ and $\eta_H$, are given by
				\begin{equation}
					\label{slowrollpa}
					\epsilon_H=-\frac{\dot H}{H^2},\qquad \eta_H=-\frac{\ddot H}{2\dot H H}.
				\end{equation}
				To guarantee an accelerated expansion and a slow-roll inflation, these parameters must meet the conditions $\epsilon_H\ll1$, $\eta_H\ll1$. The number of e-folds corresponding to this period is
				\begin{equation}
					N=\ln{\left(\frac{a_{end}}{a_i}\right)},
				\end{equation}
				where $a_{end}$ is the value of $a$ at the end of slow-roll inflation and $a_i$ its value at the onset. A period of early inflationary expansion consistent with observations demands $ N \ge 60$ \cite{wmap}.
				\item For the scalar field, the continuity equation and Klein-Gordon equation take the usual form,
				\begin{eqnarray}
					\ddot\phi+3H\dot\phi+\partial_\phi\mathcal{V}(\phi)&=0,\\
					\dot\rho+3H(\rho+P)&=0.
				\end{eqnarray}
				The modified Klein-Gordon equation for $\phi(t)$ and the corresponding continuity equation are given by
				\begin{eqnarray}
					\label{KGnc}
					\ddot\phi+3H^{nc}\dot\phi+\partial_\phi\mathcal{V}(\phi)&=0,\\
					\dot\rho+3H^{nc}(\rho+p)&=0,
				\end{eqnarray}
				where we define the the noncommutative Hubble parameter as
				\begin{equation}
					\label{Hnc}
					H^{nc}\equiv\frac{\dot{V}^{nc}}{3V^{nc}}.
				\end{equation}
				\item The modified noncommutative Friedmann equation (\cite{ijmpd}) is
				\begin{equation}
					\label{Fnc}
					H^2=\frac{8\pi G}{3}\rho\left[1-\frac{\rho+\rho_\theta}{\rho_c}\right],
				\end{equation}
				where $\rho_\theta=\rho_1(\theta)+\rho_2(\theta^2)$, and these are given as
				\begin{eqnarray}
					\rho_1(\theta)&=&\frac{2a\theta\phi}{V}(\rho_c-\rho)\nonumber\\
					&&+\frac{\sqrt 2b\theta}{\gamma\lambda V}\sqrt{\frac{\rho_{c}}{\rho}\left(1-\frac{\rho}{\rho_c}\right)(\rho-\mathcal{V}(\phi))},\\
					\nonumber
					\rho_2(\theta^2)&=&\frac{a^2\theta^2\phi^2}{V^2}(\rho_c-\rho)\nonumber\\
					&&+\frac{\sqrt 2ab\theta^2\phi}{\gamma\lambda V}\sqrt{\frac{\rho_{c}}{\rho}\left(1-\frac{\rho}{\rho_c}\right)(\rho-\mathcal{V}(\phi))}\nonumber\\
					&&+\frac{4\pi Gb^2\theta^2}{3V^2}\rho_c\left(1-\frac{\mathcal{V}(\phi)}{\rho}\right).
				\end{eqnarray}
				The complexity of the form of the modified Friedmann equation makes it difficult for us to obtain an analytical form of the condition (\ref{slowrollpa}). In this case we will carry out a numerical analysis in order to see the effects and properties of noncommutativity in the inflationary model of LQC.
				
			\end{itemize}
			
			With these defined quantities we are in a position to perform an analysis of the pre-inflationary and inflationary dynamics. As we saw above, there are two initial conditions that guarantee that the noncommutative model has a bounce with the same characteristics of standard LQC. In the following, we will discuss how these conditions affect the inflationary dynamics.
			
			
			\subsection{$b=0$, $V^{nc}(t_b)>0$}
			
			In this section we will present, with the help of some plots, the behavior of the evolution of the universe for the different periods EKED, KED, PED, after the bounce, considering condition $b=0$ (for particular initial conditions). 
			
			Taking the Hamiltonian (\ref{ncham}) specialized to the quadratic potential term, we get the equations of motion,
			\begin{equation}
				\mathcal{H}^{nc}_{\mathrm{eff}}=-\frac{3}{8\pi G\gamma^{2}\lambda^{2}}\sin^{2}(\lambda\beta)V^{nc}+\frac{(p^{nc}_{\phi})^2}{2V^{nc}}+\frac{1}{2}m^2\phi^2V^{nc}\label{ham-const}
			\end{equation}
			\begin{eqnarray}
				\label{betanc}\dot{\beta}&=&-\frac{3}{\gamma\lambda^2}\sin^{2}(\lambda\beta)+4\pi G\gamma m^{2}\phi^{2},\\
				\label{volnc}\dot{V}&=&\frac{3}{\gamma\lambda}V^{nc}\sin(\lambda\beta)\cos(\lambda\beta)-4\pi G\gamma b\theta\frac{p_{\phi}^{nc}}{V^{nc}},\\
				\label{finc}\dot{\phi}&=&\frac{p^{nc}_{\phi}}{V^{nc}},\\
				\label{pfinc}\dot{p}_{\phi}&=&\frac{3a\theta}{8\pi G\gamma^{2}\lambda^{2}}\sin^{2}(\lambda\beta)+a\theta\frac{(p_{\phi}^{nc})^{2}}{2(V^{nc})^{2}}-m^{2}V^{nc}\phi\nonumber\\
				&&-\frac{1}{2}a\theta m^{2}\phi^{2},
			\end{eqnarray}
			of course, these field equations reduce to the commutative ones when taking $\theta\rightarrow0$. Some numerical solutions for this case are given in \cite{prd}. 
			
			\subsubsection{Extreme Kinetic Energy Domination, after the bounce}
			
\begin{figure}[htp]
				\begin{center}
					\includegraphics[width=3in]{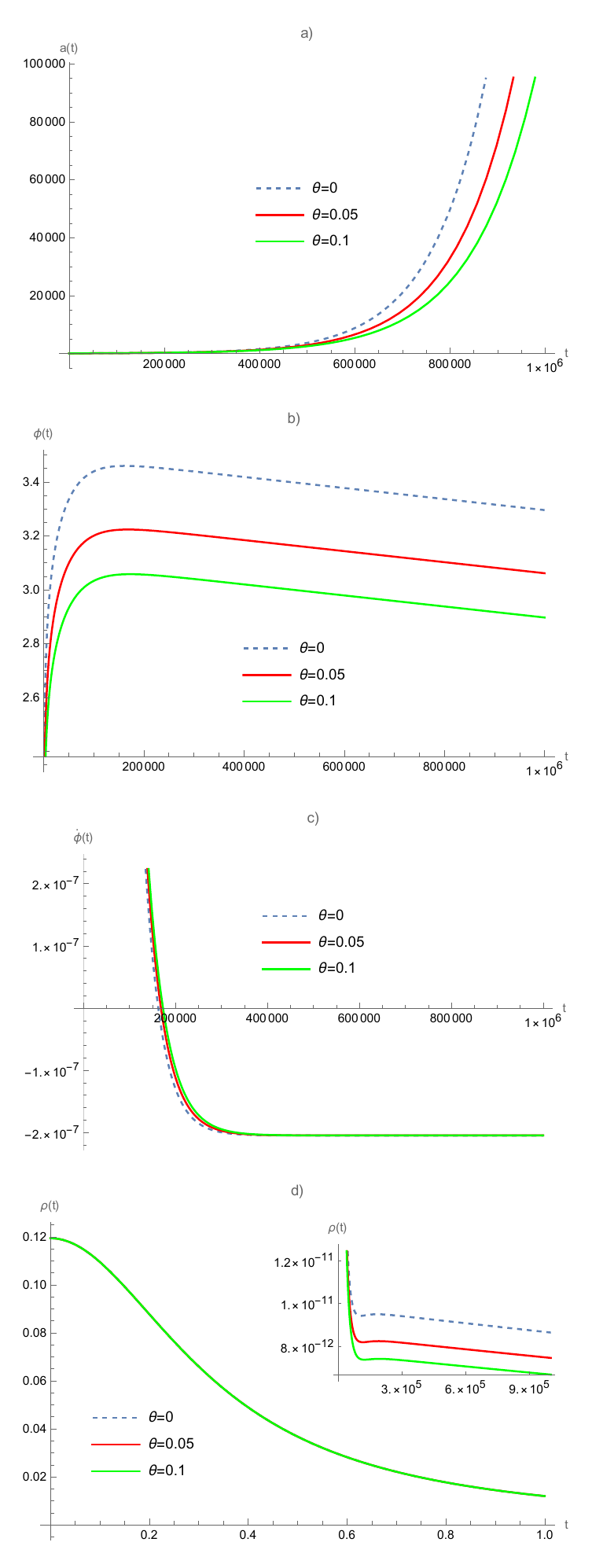}
					\caption{\justifying{Time evolution of several quantities in the EKED period of the commutative and noncommutative effective model of LQC. The initial conditions for the numerical results are $\phi_B=0$ $\dot\phi_B=0.8$, $F_B=0$ so we are in an EKED after the bounce, $m=1.26\times10^{-6}\rm{mpl}$, $b=0$. The graphs represent the commutative evolution $\theta=0$, and the noncommutative evolution $\theta=0.05$ and $\theta=0.1$.}}
					\label{fig:b0}
				\end{center}
			\end{figure}

			We will start the analysis with the initial conditions for an EKED after the bounce, $F_B<10^{-4}$, $\phi(B)=0$ and $\dot\phi(B)=0.8$. Analytical solutions to the equations of motion (\ref{betanc}-\ref{pfinc}) could not be found. However, some curves constructed via numerical solutions are plotted. The main results are shown in Figs. \ref{fig:b0}. The scale factor has an exponential growth, whereas at higher values of the noncommutative parameter, $\theta$, the growth of the scale factor is lower. On the other hand, the energy density at early times is not affected by the noncommutativity, but at larger times there is a lag in the energy density, particularly, noncommutativity accelerates the decay of the energy density with respect to time, as shown in d) of Fig (\ref{fig:b0}).

\begin{figure}[htp]
				\begin{center}
					\includegraphics[width=3in]{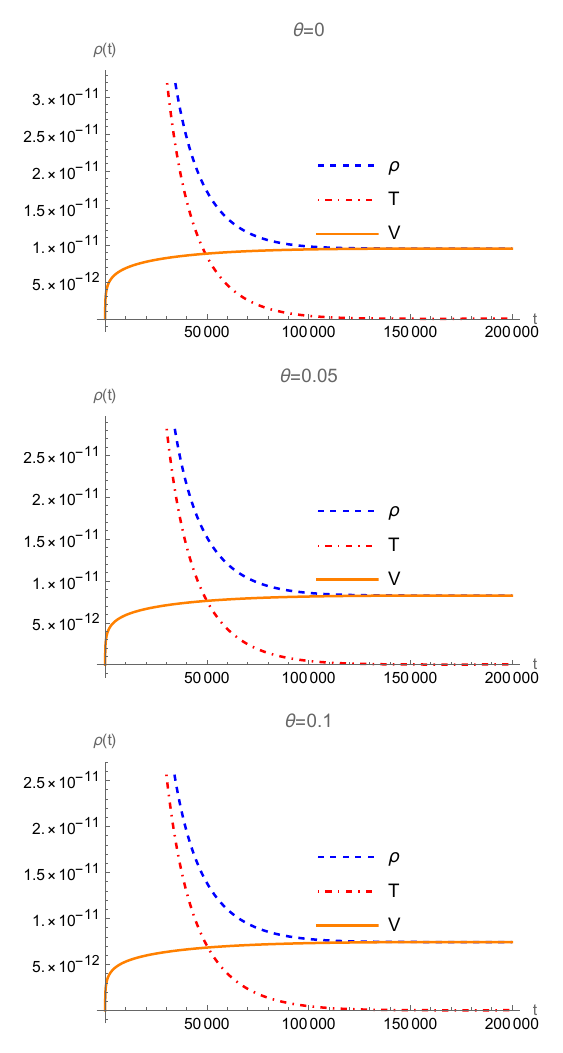}
					\caption{\justifying{The figure shows the behavior of the energy density $\rho$, the kinetic energy $T$ and the potential energy $V$, of the commutative and noncommutative models of LQC for the EKED domain. We take the same conditions as in Fig \ref{fig:b0}}}
					\label{fig:enerb0}
				\end{center}
			\end{figure}

\begin{figure}[htb]
				\begin{center}
					\includegraphics[width=3in]{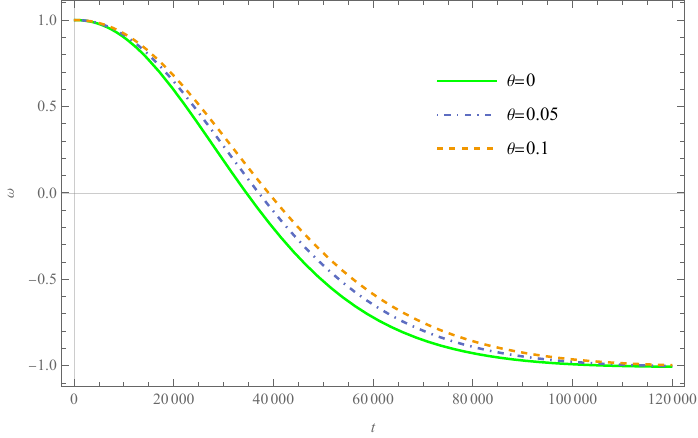}
					\caption{\justifying{Evolution of the equation of state $\omega_\phi$ for the domain of EKED for the commutative and noncommutative models. We take the same conditions as in Fig \ref{fig:b0}}}
					\label{fig:omegab0}
				\end{center}
			\end{figure}

			Fig. \ref{fig:enerb0} shows the behavior of the total, kinetic and potential energy densities for the noncommutative and commutative cases, here the domination periods of kinetic and potential energy are clearly exemplified, given the initial conditions. 
			As previously pointed out, for the EKED evolution the scalar field takes values such that $\mathcal{V}(\phi)=0$ and the Hubble parameter $H$ vanishes. In parts b) and c) of Figure \ref{fig:b0} we can see the behavior of the inflaton field. The Klein-Gordon equation tells us that, for the given the initial condition, the inflaton will slow down, $\ddot\phi< 0$, which is due to the friction term $-3H\dot\phi$, causing a loss of kinetic energy and a gain of potential energy. These behaviors can be observed in Figure \ref{fig:enerb0}. By incorporating noncommutativity into the model, we analyze the form of the modified Klein-Gordon equation (\ref{KGnc}). The behavior will be similar due to the form of Eq. (\ref{KGnc}), the term that will carry the noncommutativity correction will be the friction term $-3H_{nc}\dot\phi$. These corrections are shown in Figure \ref{fig:enerb0}. For the EKED at the bounce, the dynamics of the universe can be divided for analysis into stages. In order to have a more detailed picture of the energy domination periods, we focus on the equation of state parameter, $\omega_\phi$. Figure \ref{fig:omegab0} shows the behavior for the commutative and noncommutative scenarios. From Figure \ref{fig:omegab0} we identify three stages, the kinetic energy domination, the bounce period, 
			the transition period, and the inflationary period, similar to \cite{Parampreet}, which correspond to the pre-inflationary and inflationary dynamics of the universe. A first analysis of Fig. \ref{fig:omegab0} shows that the effect of noncommutativity is to extend the bouncing, transition, and slow-roll inflation stages, but a more detailed analysis is necessary. Next will give a more elaborated discussion of these results for each of these stages.
			%
				%
				\begin{table*}[t]
					\begin{center}
						\begin{tabular}{c   l  c  c  c  c  c c}
							\hline
							$b=0$ & Event & $t$ & $\phi$ & $\dot\phi$ &$H$ & $\dot H$ & $N$ \\ \hline
							\hline
							& Bounce & 0 & 0 & 0.8  & 0 & 0& 0\\
							& End SI &0.333 & 0.235 & 0.566 & 0.5 &$-9.185\times10^{-17}$ & 0.115  \\
							$\theta=0$& KE=PE & $3.457\times 10^{4}$ & 3.250 &$6.844\times10^{-6}$ & $1.185\times10^{-5}$ & $-2.107\times10^{-10}$& 3.886 \\
							& Onset Slow Roll & $1.075\times10^5$ & 3.444 &$6.947\times10^{-7}$ & $8.882\times10^{-6}$ &$-4.643\times10^{-17}$& 4.574 \\  
							& End of Inflation & $1.629\times10^7$ &0.2013347 &$-1.793801\times10^{-7}$ & $6.358859\times10^{-7}$& $-4.043509\times10^{-13}$& 76.594\\ \hline
							& Bounce & 0 & 0 & 0.8  & 0 & 0 &0\\
							& End SI & 0.338 & 0.232 & 0.537 & 0.529 & $-2.687\times10^{-16}$ &  0.123  \\
							$\theta=0.05$\ & KE=PE & $3.704\times10^{4}$ & 3.034 & $5.872\times10^{-6}$ & $1.106\times10^{-5}$ & $-1.836\times10^{-10}$ & 3.955\\
							& Onset Slow Roll & $1.166\times10^5$ & 3.212 &$5.463\times10^{-7}$ & $8.282\times10^{-6}$ &$-2.399\times10^{-16}$& 4.654 \\ 
							& End of Inflation & $1.514\times10^7$ &0.2013346 &$-1.793800\times10^{-7}$ & $6.358856\times10^{-7}$& $-4.043505\times10^{-13}$& 66.599\\ \hline
							& Bounce & 0 & 0 & 0.8  & 0 &0&0\\
							& End SI & 0.340 & 0.228 & 0.515 & 0.557 & $1.862\times10^{-16}$& 0.129  \\
							$\theta=0.1$\ & KE=PE & $3.900\times10^4$ & 2.881 & $5.240\times10^{-6}$ & $1.051\times10^{-5}$ &$-1.656\times10^{-10}$& 4.010 \\
							& Onset Slow Roll & $1.240\times10^5$ & 3.048 &$4.505\times10^{-7}$ & $7.861\times10^{-6}$ &$-2.975\times10^{-16}$& 4.718 \\ 
							& End of Inflation & $1.434\times10^7$ &0.2013348 &$-1.793802\times10^{-7}$ & $6.358861\times10^{-7}$&$ -4.043512\times10^{-13}$& 59.995\\ \hline
							\hline
							& Bounce & 0 & 0.917 & 0.8  & $6.123\times10^{-17}$ & 0& 0\\
							& End SI &0.333 & 1.152 & 0.566 & 0.5 &$-4.249\times10^{-16}$ & 0.115  \\
							$\theta=0$& KE=PE & $2.744\times 10^{4}$ & 4.105 &$8.609\times10^{-6}$ & $1.497\times10^{-5}$ & $-3.362\times10^{-10}$& 3.810 \\
							& Onset Slow Roll & $9.170\times10^4$ & 4.306 &$7.095\times10^{-7}$ & $1.110\times10^{-5}$ &$3.159\times10^{-17}$& 4.567 \\  
							& End of Inflation & $2.044\times10^7$ &0.2013346 &$-1.793800\times10^{-7}$ & $6.358856\times10^{-7}$&$ -4.043506\times10^{-13}$& 118.649\\ \hline
							& Bounce & 0 & 0.917 & 0.8  & $2.811\times10^{-17}$ & 0 &0\\
							& End SI & 0.303 & 1.217 & 0.442 & 0.337 & $1.432\times10^{-17}$ &  0.133  \\
							$\theta=0.05$\ & KE=PE & $2.862\times10^{4}$ & 3.937 & $7.735\times10^{-6}$ & $1.436\times10^{-5}$ & $-3.092\times10^{-10}$ & 3.956\\
							& Onset Slow Roll & $9.680\times10^4$ & 4.125 &$5.901\times10^{-7}$ & $1.064\times10^{-5}$ &$-3.392\times10^{-16}$& 4.725 \\ 
							& End of Inflation & $1.955\times10^7$ &$0.2013348$ &$-1.793802\times10^{-7}$ & $6.358862\times10^{-7}$&$ -4.043512\times10^{-13}$& 108.846\\ \hline
							& Bounce & 0 & 0.917 & 0.8  & 0 &0&0\\
							& End SI & 0.281 & 1.317 & 0.297 & 0.237 & 0& 0.145  \\
							$\theta=0.1$\ & KE=PE & $2.922\times10^4$ & 3.857 & $7.338\times10^{-6}$ & $1.406\times10^{-5}$ &$-2.967\times10^{-10}$& 4.057 \\
							& Onset Slow Roll & $9.945\times10^4$ & 4.039 &$5.363\times10^{-7}$ & $1.041\times10^{-5}$ &$-6.402\times10^{-16}$& 4.833 \\ 
							& End of Inflation & $1.913\times10^7$ &$0.2013347$ &$-1.793800\times10^{-7}$ & $6.358858\times10^{-7}$&$ -4.043508\times10^{-13}$& 104.330\\ \hline
						\end{tabular}
						\caption{The table shows the numerical results of the commutative and noncommutative models of LQC with the potential $\phi^2$. The Events: bounce, the end of SI (superinflation), the equilibrium point KE=PE, the onset of slow-ll inflation, and the end of inflation, are compared for $\theta=0$, $\theta=0.05$ and $\theta=0.1$.}
						\label{tab:b0EKE}
					\end{center}
				\end{table*}
			\begin{itemize}				
				\item {\it The bouncing stage EKED} \\
				The beginning of this stage is at the bounce $t_B^{EKED}$ and ends at the time $t_{eq}^{EKED}$ when the kinetic energy of the scalar field is equal to the potential energy, regarding the equation of state parameter (\ref{omegastate}), this occurs when $\omega_\phi=0$. At the end of the bouncing stage, the LQG-related corrections can be neglected and the classical Friedmann dynamics can be recovered. For the initial conditions $(\phi_B=0,~\dot\phi_B=0.8)$ the end of the bouncing stage is at $t_{eq}^{EKED}=3.457\times10^4$ as shown in Table \ref{tab:b0EKE}. At this stage there is a period where the Hubble parameter grows very rapidly, which we call super inflation (SI). This SI period lasts for fractions of seconds. In a) of Fig. \ref{fig:SIb0} the behavior of the Hubble parameter in the SI period is shown. On the other hand, in b) of Fig. \ref{fig:SIb0} we can see that the SI period comes to an end ($\dot H=0$) at $t_{SI}^{EKED}=0.333$, and it lasts for $N(t_{SI}^{EKED})=0.115$ e-folds, for the commutative case, Table \ref{tab:b0EKE}. These results are similar to those reported in \cite{sloan} for EKED. 
				
				Noncommutativity effects in this stage vary depending on the evolution of the scalar field. In SI, Fig. \ref{fig:SIb0}, the noncommutativity effects cause the duration of SI to be larger and reach a larger growth of the commutative Hubble parameter $H(t_{SI}^{EKED})=0.5$, for $\theta=0.05$ and $\theta=0.01$ the end of SI occurs at $t_{SI_{,0.05}}^{EKED}=0.338$ and $t_{SI_{,0.1}}^{EKED}=0.340$ respectively, while the values of maxima reached by $H$ are $H(t_{SI_{,0.05}}^{EKED})=0.529$ and $H(t_{SI_{,0.1}}^{EKED})=0.557$. On the other hand, from the SI term, the noncommutativity effects extend the bouncing stage, for example the bouncing stage term occurs at $t_{eq_{,0.05}}^{EKED}=3.704\times10^{4}$ for $\theta=0.05$, while which for $\theta=0.1$ $t_{eq_{,0.1}}^{EKED}=3.900\times10^4$. These data are reported in Table \ref{tab:b0EKE}.
				\begin{figure}[htp]
					\begin{center}
						\includegraphics[width=3in]{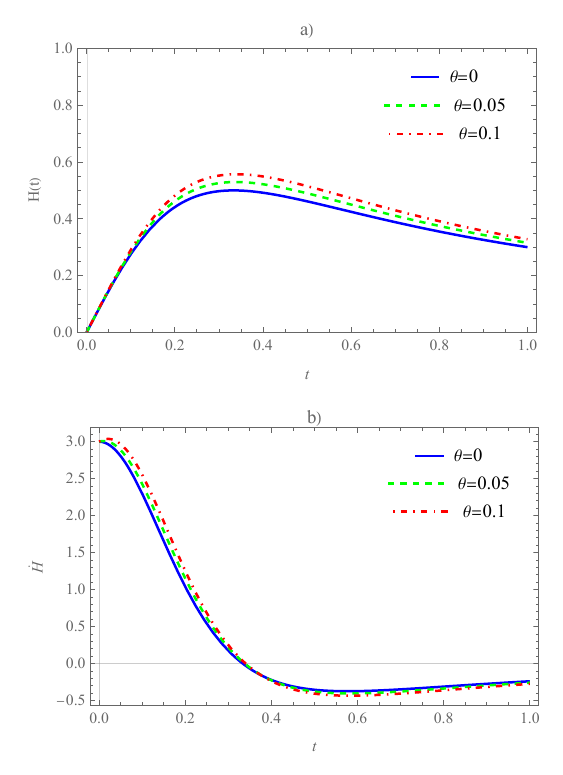}
						\caption{\justifying{The figure represents in a) the graph of the Hubble parameter $H(t)$, in b) the graph of $\dot H(t)$ for EKED of the commutative and noncommutative LQC models. We take the same conditions as in Fig \ref{fig:b0}}}
						\label{fig:SIb0}
					\end{center}
				\end{figure}
				As already pointed out, at this stage, LQG-related quantum corrections are present, so the primordial power spectra are affected by them. In this respect, several investigations have been conducted (see, e.g., section 4.2 of the first review in \cite{reviews-lqg}). Two specific frameworks have been mostly employed, the so called dressed-metric approach (see, for instance, \cite{dressed-metric} and references therein) and the hybrid quantization scheme (a recent review on the latter is given in Ref. \cite{hybrid-quant}; in \cite{dressed-metric-hybrid-quant} a relation between the two approaches at the effective level is established). As we have seen, noncommutativity effects appear to be present in a rather simple form at this stage (essentially extending its time duration). We therefore stress that, within the scope of the present analysis, the universe predicted by the $\theta-$deformed effective model put forward presently does not deviates significantly from the universe predicted by standard (effective) LQC. Put another way, for appropriate values of the $\theta$ parameter, both models are essentially indistinguishable. This conclusion had been anticipated in the preliminary investigation \cite{prd}, and the present analysis puts such conclusion on rather firmer grounds. However, a focused study on the primordial perturbations (which are not included here) may prove to be interesting and seems to be a natural complement to the present analysis: Noncommutativity may influence the universe in a more subtle manner at this more fundamental level, and tuning of the $\theta$ parameter could be exploited with an eye on taming power spectrum-related tensions still prevalent on standard LQC. Such a study lies outside the scope of the present investigation but it is currently being pursued by the authors within the hybrid quantization scheme, and will be reported elsewhere.
				%
				\item {\it Transition stage EKED}\\
				In the transition period the kinetic energy decreases abruptly, which causes the universe to accelerate, while $\omega_\phi^{EKED}$ tends to behave in a constant manner. This stage starts at equal energies $\omega_\phi^{EKED}=0$ and ends when $\omega_\phi^{EKED}=-1$ (see Fig. \ref{fig:omegab0}). For the commutative case, the stage starts at $t_{eq}=3.457\times10^{4}$ and ends at $t_{i}=1.075\times10^{5}$, where $t_{i}$ will also be the time at which slow-roll inflation begins. In terms of e-folds, this period is shorter in duration than the bouncing and slow-roll inflation stages, this kind of behavior is also reported in \cite{piulqc}. The noncommutativity effects in this stage remain the same as those at the end of the bouncing stage, at larger values of the noncommutative parameter $\theta$, the duration of the transition stage is prolonged, for example, in the commutative case,we have $N_{tr}=0.688$, while for the noncommutative cases $\theta=0.05$ and $\theta=0.1$ are $N_{tr_{0.05}}=0.699$ and $N_{tr_{0.1 }}=0.708$, respectively (Table \ref{tab:b0EKE}).
				\item {\it Slow-roll inflation EKED}\\
				When $\omega_\phi=-1$ the slow-roll inflation stage begins. At this point the energy density is totally dominated by the potential $\phi^2$, and a time $t_i=1.075\times10^5$ has elapsed , the corresponding value of the field is $\phi(t_i)=3.444$. These results are similar to those reported in \cite{sloan, Parampreet}. Figure \ref{fig:epsilonb0} shows the regions in which inflationary periods are found, $\epsilon_H\ll1$. This stage continues until the value of $\epsilon_H=1$ is reached, which occurs at $t_{end}=1.629\times10^{7}$ and $N=76.594$. A period of early inflationary expansion consistent with observations demands $ N \ge 60$ \cite{wmap}, the results are shown in Figure \ref{fig:efoldb0} and Table \ref{tab:b0EKE}.These results indicate that for EKED after the bounce, there is an inflationary period with more than 60 e-folds for the given initial conditions.
				\begin{figure}[htb]
					\begin{center}
						\includegraphics[width=3in]{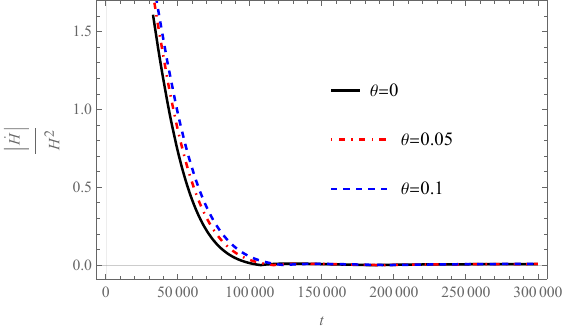}
						\caption{\justifying{Plot of the commutative and noncommutative slow-roll parameter $\epsilon(t)$ for the EKED period. We take the same conditions as in Fig \ref{fig:b0}}}
						\label{fig:epsilonb0}
					\end{center}
				\end{figure}
				In the noncommutative scenario, slow-roll inflationary dynamics behave in a similar way to its commutative counterpart. However, the effects of noncommutativity imply that the duration of the inflationary period is shorter when the parameter $\theta$ increases in value. We can see this by calculating the number of e-folds as a function of $t$,
				\begin{equation}
					\label{Nefold}
					N(t)=\int_{0}^{t}H(\tau)d\tau,
				\end{equation}
				where $H(\tau)$ is given by (\ref{Fnc}). Fig. \ref{fig:efoldb0} shows the behavior of $N(t)$. As shown in Table \ref{tab:b0EKE}, the corresponding number of e-folds for $\theta=0.05,0.1$ are $N_{.05}=66.599$ and $N_{.1}=59.995$, respectively, being less in number than those obtained in the commutative counterpart. Given this behavior, large values of $\theta$ imply that the duration of the inflationary period does not reach the minimum number of e-folds required for the universe to evolve up to the present epoch, this is one of the reasons why we do not notice the noncommutative effects, for the universe to evolve in the proper way it is necessary that $\theta\ll1$.
				\begin{figure}[htb]
					\begin{center}
						\includegraphics[width=3in]{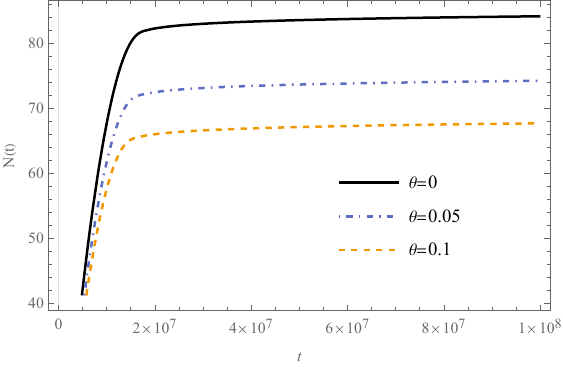}
						\caption{\justifying{Plot of the commutative and noncommutative e-folds $N(t)$ for the EKED period. We take the same conditions as in Fig \ref{fig:b0}}}
						\label{fig:efoldb0}
					\end{center}
				\end{figure}
			\end{itemize}
			
\begin{figure*}[htb]
				\begin{center}
					\includegraphics[width=6in]{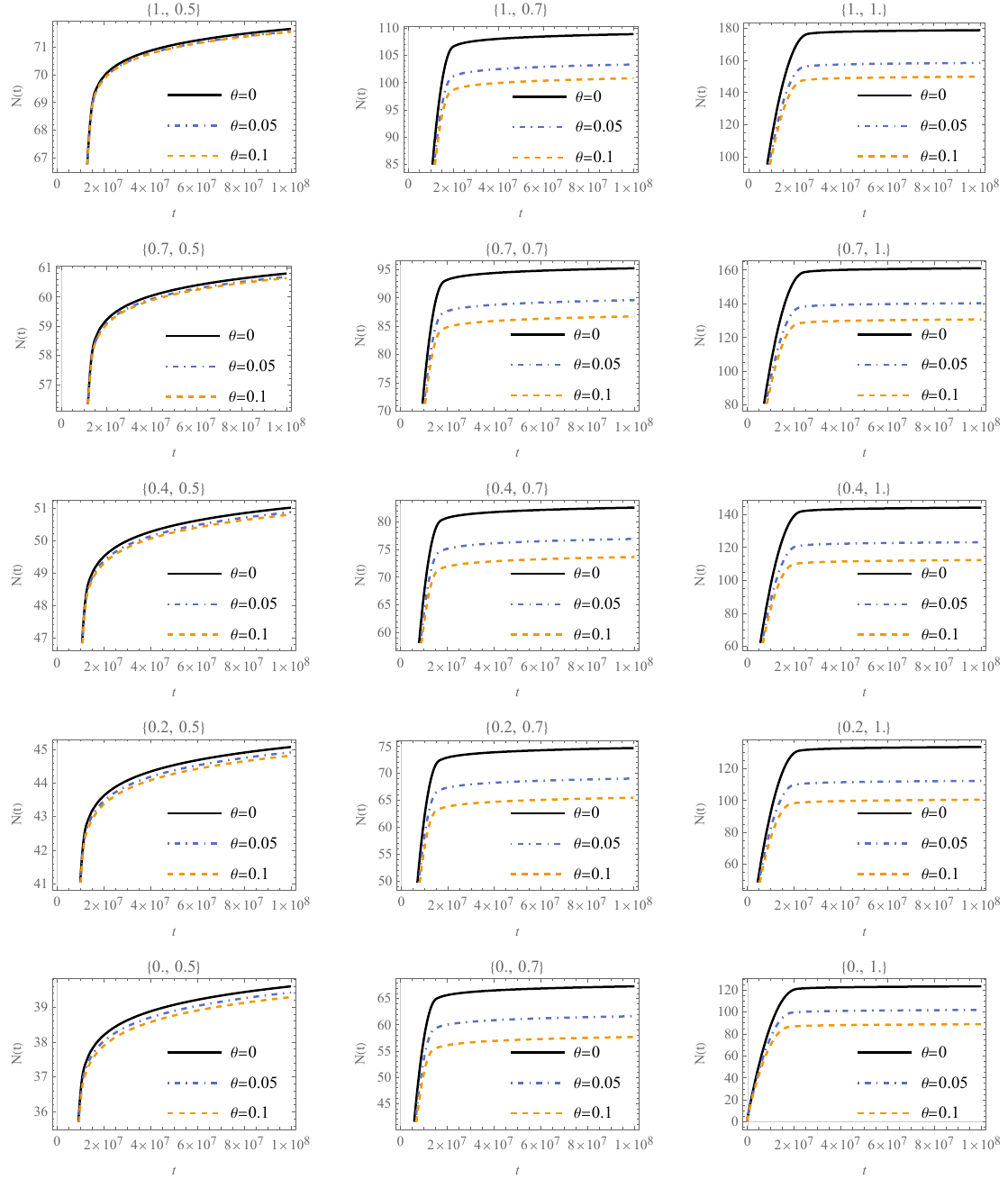}
					\caption{\justifying{The Figure shows the commutative and noncommutative $N(t)$, for different initial conditions $(\phi_B,\dot\phi_B)$. Each of the columns keeps $\dot\phi_B$ constant and for each row $\phi_B$ vary. We take values of $\dot\phi_\ge0.5$.}}
					\label{fig:multefold2}
				\end{center}
			\end{figure*}

\begin{figure*}[t]
				\begin{center}
					\includegraphics[width=6in]{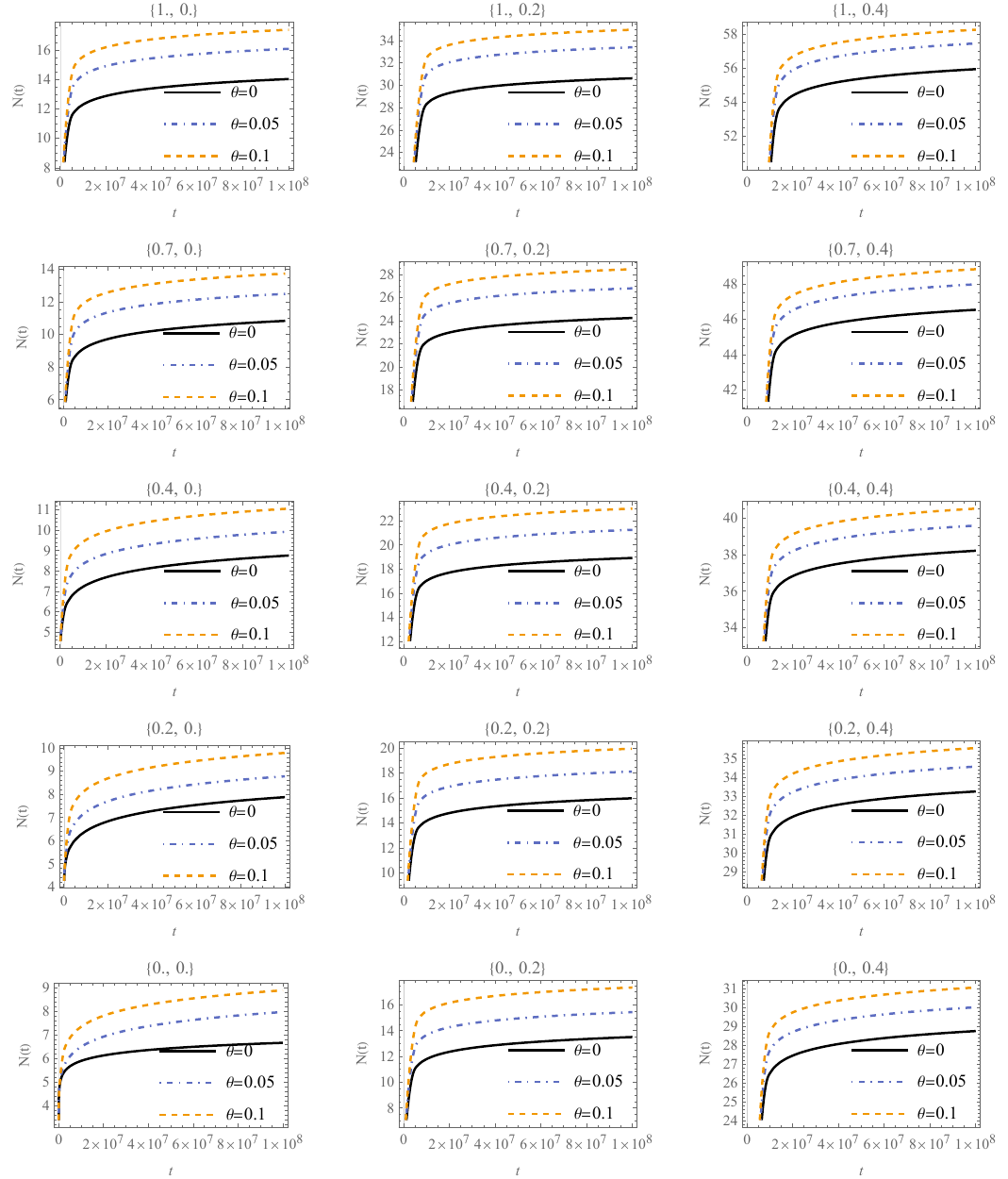}
					\caption{\justifying{The Figure shows the commutative and noncommutative $N(t)$, for different initial conditions $(\phi_B,\dot\phi_B)$. Each of the columns keeps $\dot\phi_B$ constant and for each row $\phi_B$ vary. We take values of $\dot\phi_<0.5$.}}
					\label{fig:multefold1}
				\end{center}
			\end{figure*}

			
			The behavior described above for $N(t)$ holds for the initial conditions $(\phi_B>0, ~ \dot\phi_B>=0.5)$, Fig. \ref{fig:multefold2} shows some examples with initial conditions in this range. A peculiar behavior of the noncommutative model is that for initial conditions $(\phi_B>0, ~ \dot\phi_B<0.5)$, the number of e-folds, $N(t)$, increases as the parameter $\theta$ increases. Therefore, in this scenario, it could very well be the case that commutative standard models with initial conditions not consistent with long enough inflationary periods could, upon introduction of noncommutativity, transit to models that can accommodate for a sufficiently large number of e-folds consistent with cosmological observations (see Fig. \ref{fig:multefold1}). In standard LQC, theoretical studies on global aspects of the space of solutions suggest that these initial conditions not consistent with a sufficiently long inflationary period are otherwise rare \cite{sloan}. Assuming, nonetheless, that the universe evolved from such uncommon, from the point of view of LQC, initial conditions could help to find an upper bound for the noncommutative parameter.
			%
			
			\subsubsection{Kinetic Energy Domination, after the bounce}
\begin{figure}[htb]
				\begin{center}
					\includegraphics[width=3in]{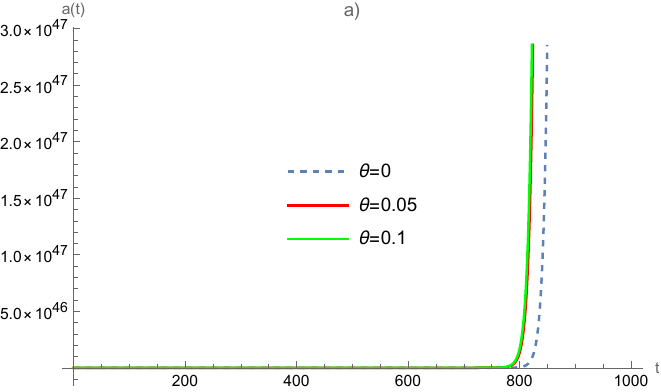}
					\caption{\justifying{Numerical solution for the scale factor evolution $a(t)$, considering the commutative and noncommutative model with the potential $\phi^2$, for the initial conditions $\phi_B=5\times10^4$, $\dot\ phi_B=0.8$, $m=1.26\time10^{-6}$, $b=0$, in KED after the bounce.}}
					\label{fig:akd}
				\end{center}
			\end{figure}

\begin{figure}[htb]
				\begin{center}
					\includegraphics[width=3in]{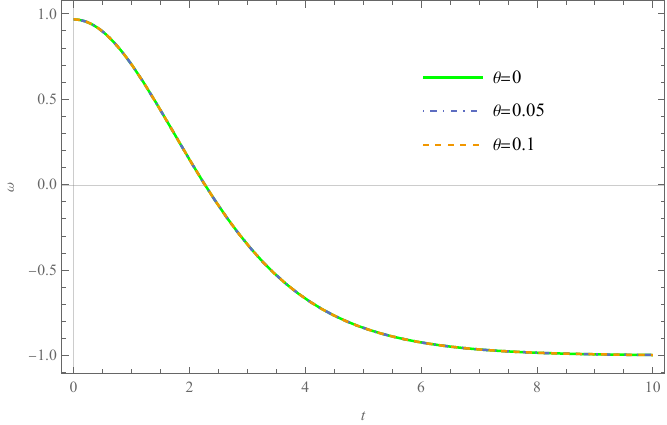}
					\caption{\justifying{Numerical solution for the state equation evolution $\omega_\phi$,in KED after the bounce. We take the same initial condition as the Fig \ref{fig:akd}.}}
					\label{fig:omegakd}
				\end{center}
			\end{figure}

			For an KED after the bounce $10^{-4}<F_B<0.5$, we take $\phi(B)=5\times10^4$ and $\dot\phi(B)=0.8$, $F_B=1.662\times10^{-2}$. The commutative behavior of the solutions for the initial conditions is similar to the case of EKED, there is a bounce at $t_B=0$ and an exponential growth in the scale factor (see Fig. \ref{fig:akd}). In contrast to the EKED case, the larger the values taken by the noncommutative parameter, $\theta$, the faster $a(t)$ grows.
			In this KED period, the bouncing stage, the transition stage and the slow-roll inflation stage are still featured, although the duration of these stages is considerably modified. In Fig.~\ref{fig:omegakd} we can see this behavior, as well as the variation between the commutative and noncommutative parts, in the $\omega_\phi$, is minimal.
			\begin{itemize}
				\item {\it The bouncing stage KED} \\ Again we find an SI stage, in which we have the growth of $H(t)$, where it reaches a maximum value of $H_{max}=0.5$. The duration of this period is $t=0.341$ with $N=0.118$, similar to that of EKED ($t=0.333$,$N=0.115$). The noncommutative model of SI presents a much shorter duration than its commutative counterpart, $t=3.503\times10^{-3}$ for $\theta=0.05$, while for $\theta=0.1$ , SI ends at $t=2.477\times10^{-3}$. Another characteristic that differentiates the EKED and KED periods from their commutative counterparts are the maximums reached by the Hubble parameter $H(t)$, while for the commutative case the difference is minimal, for the noncommutative case there are big differences. For example, in EKED $H_{0.05}^{\rm{\small EKED}}=0.529$, $H_{0.1}^{\tiny EKED}=0.557$, while in KED $H_{0.05}^{KED}=95.137$, $H_{0.1} ^{KED}=134.544$. From the point of view of the evolution of $H(t)$,  the commutative case KED behaves similarly to EKED for the maximum value of $H(t)$. Analyzing the behavior of $H(t)$, for the noncommutative case, the increase of $\theta$ causes the maximum value of $H(t)$ to increase, so the existence of $\theta$ implies that the SI stage has a larger amplitude (regarding the Hubble parameter) compared to its commutative counterpart, see Fig.~\ref{fig:Hkd}. The end of the bouncing stage occurs when KE = PE, at $\omega_\phi^{KED}=0$. In the case of KED, the commutative and noncommutative behaviors at this time are similar up to $10^{-6}$, and there are fluctuations of up to $10^{-7}$.
				\begin{figure}[htb]
					\begin{center}
						\includegraphics[width=3in]{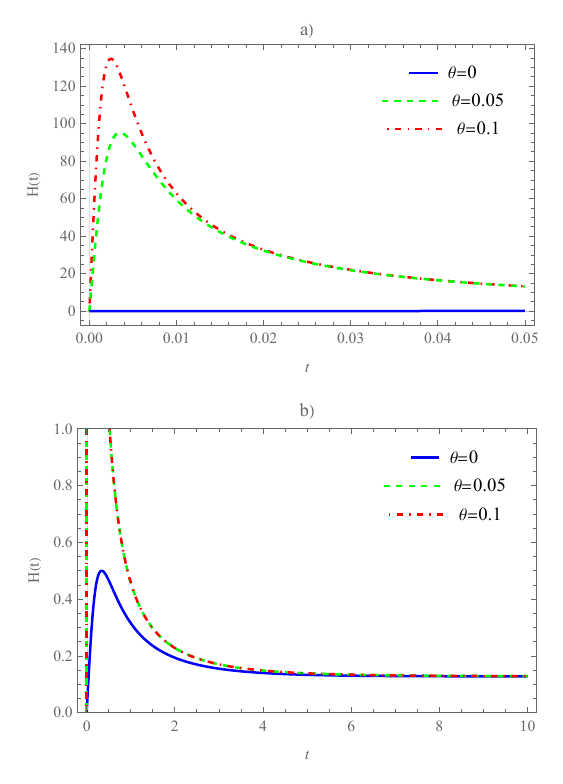}
						\caption{\justifying{Noncommutative and commutative numerical solution for the Hubble parameter evolution $H(t)$,in KED after the bounce. We take the same initial condition as the Fig \ref{fig:akd}. In part a) it is shown how the noncommutative parameter $\theta$ causes a considerable increase in the maximum value of $H(t)$ for the SI stage. b) It is shown how $H(t)$ for the commutative case has a SI stage.}}
						\label{fig:Hkd}
					\end{center}
				\end{figure}
				\item {\it Transition stage KED}\\
				In the KED period there is a much shorter transition stage than in EKED. For the commutative case, the EKED period has a duration of $\Delta t^{EKED}_{tr}=72930$, while the KED period lasts $\Delta t^{KED}_{tr}=21.3664$. For the noncommutative case there is a similar behavior. As in EKED, the kinetic energy decreases rapidly, the end of the transition stage happens when $\omega_\phi^{KED}=-1$, for the commutative case this happens at $t_{tr}^{KED}=23.6330$, while for the noncommutative cases $\theta=0.05$ and $\theta=0.1$ it happens at $t_{tr,0.05}^{KED}=23.6336$ and $t_{tr,0.1}^{KED}=23.6329$ respectively, which is consistent with Fig. \ref{fig:omegakd}.
				
				\item {\it Slow-roll inflation KED}\\
				Fig. \ref{fig:epsilonkd} shows that there is a region in which $\epsilon^{KED}<1$, which implies that the commutative and noncommutative models experience an accelerated expansion. In Fig. \ref{fig:epsilonkd} there are two regions in which $\epsilon^{KED}<1$, the first region consists of the super inflation stage, while the second region starts at $t_i^{KED}=2.919$ for the commutative case, while for the noncommutative model it starts at $t_{i,0.05}^{KED}=3.22962$ and $t_{i,0.1}^{KED}=3.22963$. At $t_i^{KED}, t_{i,\theta}^{KED}$ times, the universe begins to undergo acceleration, but also $\omega_\phi^{KED}\ne-1$, so there are still contributions from kinetic energy. To ensure that we are in slow-roll inflation it is necessary to review the second parameter of slow-roll.
				\begin{figure}[htb]
					\begin{center}
						\includegraphics[width=3in]{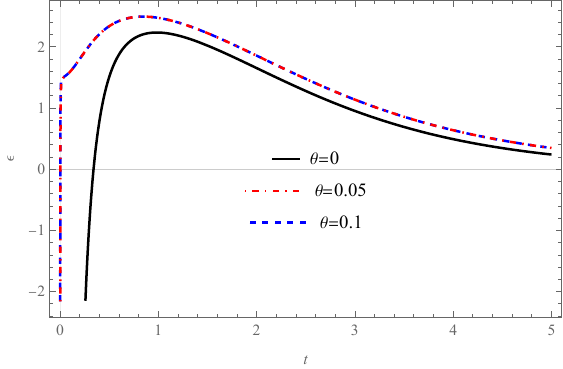}
						\caption{\justifying{Noncommutative and commutative numerical solution for the slow-roll parameter evolution $\epsilon(t)$,in KED after the bounce. We take the same initial condition as the Fig. \ref{fig:akd}.}}
						\label{fig:epsilonkd}
					\end{center}
				\end{figure}
				Fig. \ref{fig:etakd} show us the behavior of the second slow-roll parameter $|\eta|$. As in \cite{Parampreet}, the beginning of the slow-roll inflation stage is when $|\eta|<0.03$. Fig. \ref{fig:etakd} shows us that $|\eta|$ has large oscillations in its behavior, but on average its value is decreasing and tends to zero. This happens (at an average value) when $t=111.0405$, $t_{0.05}=86.6205$, $t_{0.1}=72.5000$. From here we can see that by increasing the noncommutative parameter, $\theta$, the start of the slow-roll stage occurs at earlier times.
				\begin{figure}[htb]
					\begin{center}
						\includegraphics[width=3in]{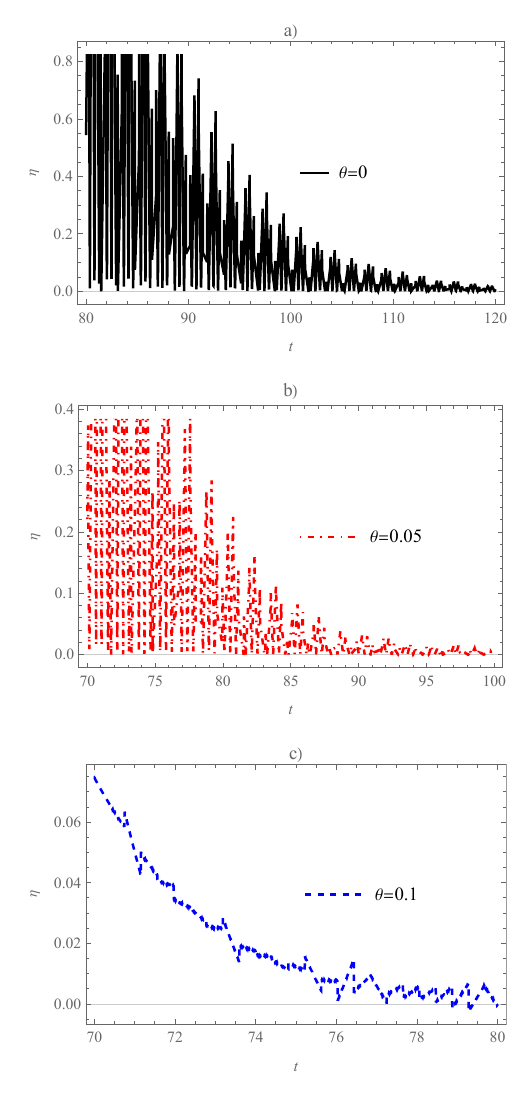}
						\caption{\justifying{Noncommutative and commutative numerical solution for the slow-roll parameter evolution $\eta(t)$,in KED after the bounce. We take the same initial condition as the Fig. \ref{fig:akd}.}}
						\label{fig:etakd}
					\end{center}
				\end{figure}
				Similar to the EKED and KED, we can compute the number of e-folds as a function of time $N(t)$. Since the PED has a larger range than the EKED and KED, the number of e-folds grows much faster in time compared to the both EKED and KED. Fig. \ref{fig:efoldkd} shows that a $t=1000$ the amount of e-folds is $128.467$ for the commutative case while $N_{0.05}=131.741$ and $N_{0.1}=131.972$ for the noncommutative counterparts. The effect, on the number of e-folds, of including noncommutativity is that the greater the value of the parameter $\theta$, the greater the number of e-folds, which contrasts with the EKED associated to condition $\dot{\phi_B}=0$.
				\begin{figure}[htb]
					\begin{center}
						\includegraphics[width=3in]{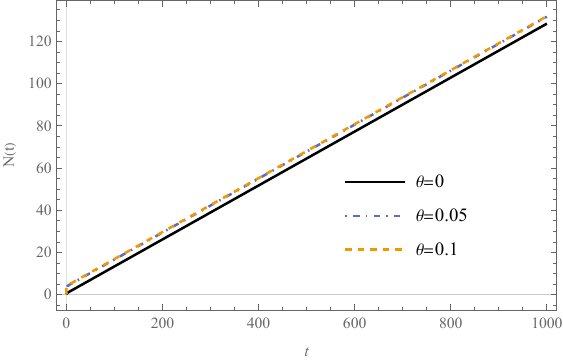}
						\caption{\justifying{Noncommutative and commutative numerical solution for $N(t)$, in KED after the bounce. We take the same initial condition as the Fig \ref{fig:akd}.}}
						\label{fig:efoldkd}
					\end{center}
				\end{figure}
				Due to the behavior of the slow-roll parameters, the universe will remain in accelerated expansion and eternal inflation, such as those presented in \cite{Guth}. 
			\end{itemize}
			
			\subsubsection{Potential Energy Domination, after the bounce}
\begin{figure}[htb]
				\begin{center}
					\includegraphics[width=3in]{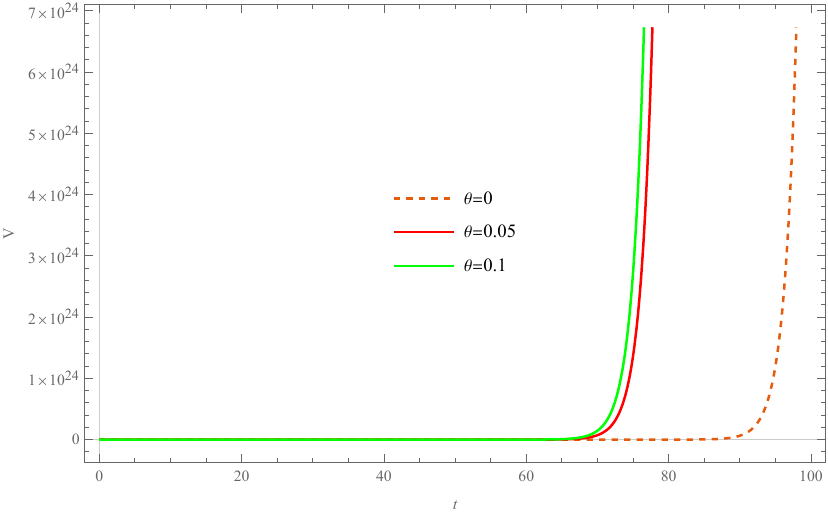}
					\caption{\justifying{Numerical solution for volume evolution $V(t)$, considering the commutative and noncommutative model with the potential $\phi^2$, for the initial conditions $\phi_B=3.8\times10^5$, $\dot\phi_B=0.1$, $m=1.26\time10^{-6}$, $b=0$, in PED after the bounce.}}
					\label{fig:volpd}
				\end{center}
			\end{figure}

\begin{figure}[htb]
				\begin{center}
					\includegraphics[width=3in]{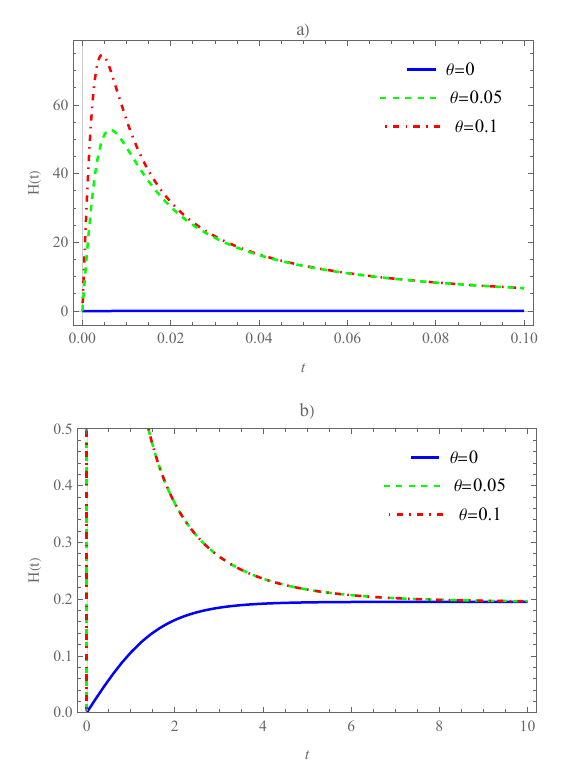}
					\caption{\justifying{Noncommutative and commutative numerical solution for $H(t)$, in PED after the bounce. We take the same initial condition as the Fig \ref{fig:volpd}.}}
					\label{fig:Hdp}
				\end{center}
			\end{figure}

\begin{figure}[htb]
				\begin{center}
					\includegraphics[width=3in]{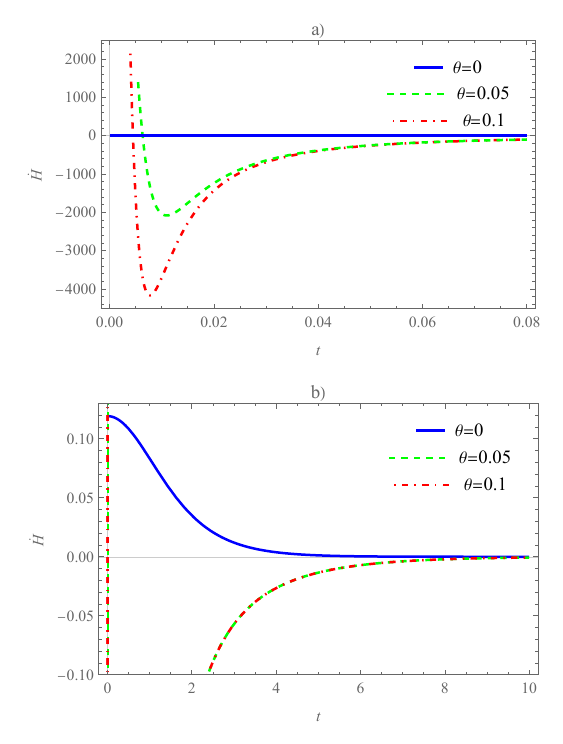}
					\caption{\justifying{Noncommutative and commutative numerical solution for $\dot H(t)$, in PED after the bounce. We take the same initial condition as the Fig \ref{fig:volpd}.}}
					\label{fig:Hpdp}
				\end{center}
			\end{figure}

\begin{figure}[htb]
				\begin{center}
					\includegraphics[width=3in]{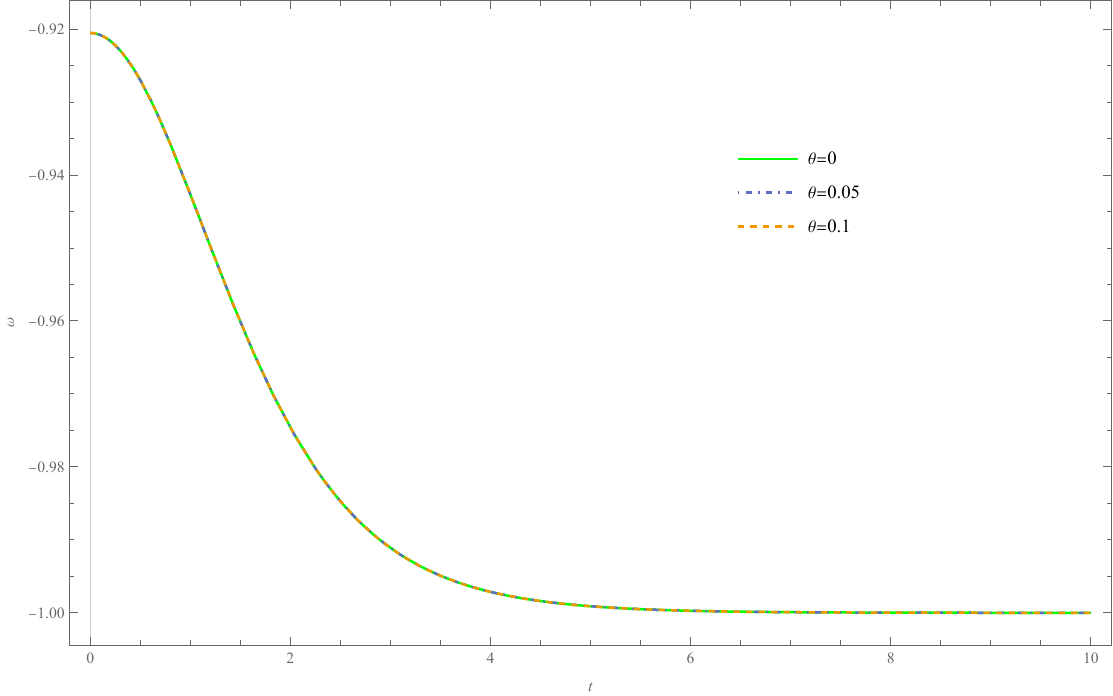}
					\caption{\justifying{Noncommutative and commutative numerical solution for $\omega_\phi$, in PED after the bounce. We take the same initial condition as the Fig \ref{fig:volpd}.}}
					\label{fig:omegapd}
				\end{center}
			\end{figure}

			In this stage the potential energy dominates the evolution of the universe after the bounce, $F_B>0.5$. For the analysis of this stage we will take the initial conditions $(\phi_B=3.5\times10^5, ~ \dot\phi_B=0.1)$, which implies that $F_B=0.9$. As the potential energy dominates, the universe is in an inflationary period (the bouncing and transition phases are not present in this stage). As in Refs. \cite{sloan, Parampreet}, we find that the universe has an exponential growth. Noncommutativity has the effect of increasing the rate to which the volume of the universe grows (with increasing values of $\theta$) see Fig. \ref{fig:volpd}. 
			The commutative Hubble parameter in the PED grows and then remains constant as the universe evolves \cite{Parampreet}. 
			On the other hand, incorporating noncommutativity in this stage produces the appearance of a very brief SI stage, as the noncommutative parameter, $\theta$, increases. The increase of the Hubble parameter, $H$, is greater, Fig. \ref{fig:Hdp}. After a short period of time it tends to the same constant value featured in the commutative counterpart.
			The duration of the SI induced by noncommutativity one is very short, for $\theta=0.05$ we have that the end of SI is $t_{0.05}=6.32\times10^{-3}$, while for $ \theta=0.1$ the end of SI occurs in less time $t_{0.1}=4.47\times10^{-3}$, as shown in Fig. \ref{fig:Hpdp}.
			From Fig. \ref{fig:omegapd} we can clearly visualize that the kinetic energy after the bounce is almost zero and that the potential energy dominates after the bounce, the bouncing and transition stages do not appear because $\omega_{\phi}\sim-1$, that is to say that we are in the slow-roll stage, this behavior is shared by both, commuative and noncommutative schemes.

			\begin{figure}[htb]
				\begin{center}
					\includegraphics[width=3in]{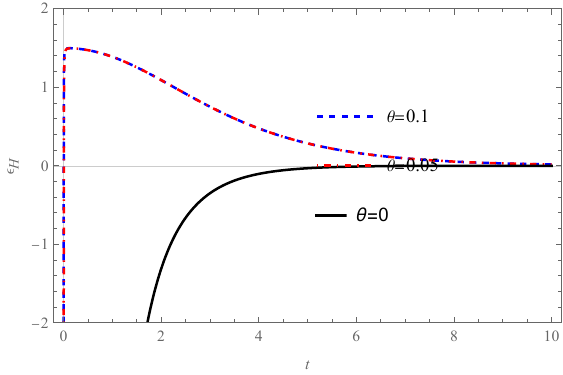}
					\caption{\justifying{Noncommutative and commutative numerical solution for $\epsilon_H(t)$, in PED after the bounce. We take the same initial condition as the Fig \ref{fig:volpd}.}}
					\label{fig:epsilondp}
				\end{center}
			\end{figure}

			\section{Discussion}\label{section4}
			In this letter we examined the pre-inflationary and inflationary epochs of the noncommutative effective LQC (with a massive scalar field with quadratic potential) presented in \cite{prd}. As in \cite{sloan}, we analyzed three periods of energy domination after the bounce, the EKED, KED and PED, which are specified by their corresponding values of $F_B$, Eq. \ref{Fb}. In \cite{prd} there are two sets of conditions that ensure the existence of a bounce, for this investigation only the condition $b=0$ was taken into account (since the other one does not precludes existence of additional minima of volume, and dramatically reduce the number of compatible initial conditions). 
			
			The noncommutativity effects are expected to have its main contribution in EKED and KED after the bounce, more precisely, at the bouncing stage of these scenarios.
			
			\subsection{Regarding the EKED}  
			
			The initial conditions compatible with the EKED period are $\phi_B$, $F_B<10^{-4}$. For this analysis we used $\phi_B=0, ~ \dot\phi_B=0.8$. By solving the field equations (\ref{betanc}-\ref{pfinc}) the numerical solutions for the commutative and noncommutative evolution of the effective LQC model with the potential $\phi^2$ were found. Using the modified Friedmann equation \ref{Hnc} and the $\omega_{\phi}$ \ref{omegastate} equation of state for the scalar field, the post bounce evolution can be divided into the bouncing stage, transition stage and the slow-roll inflation. For the commutative case $\theta=0$, the results obtained in these three stages are similar to the results reported in \cite{sloan,Parampreet,piulqc}, as expected. By taking $\theta\ne0$, the equation of state $\omega_\phi$ undergoes modifications for each value of $\theta$, which causes a displacement at the beginning of the pre-inflationary and inflationary stages. Taking increments in $\theta$, the stages occur at later times than their commutative counterparts ($\theta=0$). In the bouncing stage, the SI period is modified by an increase in the length of the period and by the maximum value reached by $H(t)$, see Tab. \ref{tab:b0EKE}. There is slow-roll inflation, which occurs when $\omega_\phi=-1$, Fig. \ref{fig:efoldb0} shows the interval in which $\epsilon_H<1$. Finding the number of e-folds $N(t)$ we observe that the minimum bound $N\ge60$ \cite{wmap} is reached for $\theta=0$. For values of $\theta>0$, $N(t)$ decreases with respect to its commutative counterpart, that is, for $\theta\gg1$ the minimum number of e-folds is not reached. Therefore, to guarantee the condition $N\ge60$, the value of the noncommutative parameter must necessarily be close to zero, 
			for the given initial conditions (this, of course, is consistent with the idea that the noncommutative parameter must be ``small''). In addition to this initial condition, several other initial conditions $\phi_B,\dot\phi_B$ were analyzed to see if there is a similar behavior for $\theta\ne0$. It was found that this is indeed the case for initial conditions where $(\phi_B,~\dot\phi_B>0.5)$. 
			But for conditions where $(\phi_B,0\le\dot\phi_B<0.5)$, $N\ge60$ is not reached for $\theta\ne0$, (not even for $\theta=0$). For these conditions the noncommutative behavior is inverted, as $\theta$ increases, now the number of noncommutative e-folds is greater than the number of e-folds 
			reached  in the commutative case, see Figs. \ref{fig:multefold2}, \ref{fig:multefold1}.
			\subsection{Regarding the KED}
			In the KED, $10^{-4}<F_B<0.5$, the pre-inflationary and inflationary stages prevail, since the potential energy density has a greater contribution compared to the EKED, both the bouncing and transition stages are reduced temporarily, while slow-roll inflation exists in earlier times. The noncommutative effects do not significantly affect the duration of the stages, see Fig. \ref{fig:omegakd}, the fluctuations due to noncommutative effects are in the order of $10^{-7}$. The main effect of noncommutativity in KED is in the SI period, as in EKED, the effect of noncommutativity increases the duration and the maximum value of $H(t)$ when $\theta$ increases. In KED, there is a much larger maximum increase of $H(t)$ than for EKED, see Fig. \ref{fig:Hkd}, which implies that in the SI period noncommutativity increases the range of growth of the scale factor in such a way that the universe grows and expands at a faster rate.
			For the end of the transition stage $\omega_\phi=-1$, but in KED the field $\ddot\phi$ still contributes to the dynamics of the system. 
			Fig. \ref{fig:efoldkd} shows the range in which $\epsilon_H< 1$, while Fig. \ref{fig:etakd} indicates the start of the slow-roll inflation stage at $t_i^ {KED}\approx111.0405$. 
			By taking $\theta\ne0$ the period in which $\epsilon_H< 1$ occurs at later times than its commutative counterpart, the opposite effect appears for the condition $\eta_H\ll1$, which for $\theta>0$, occurs earlier than the $\theta=0$ case. The condition $N\ge60$ is fulfilled for both commutative and noncommutative cases ($N(t)$ increases with $\theta$), see Fig \ref{fig:efoldkd}.
			\subsection{Regarding the PED}
			For PED, $F_B>0.5$, the potential energy density completely dominates after the bounce, the bouncing and transition stages disappear in PED. 
			For the commutative case there is an SI period, which is partially spliced with the slow-roll inflation stage, just like in \cite{Parampreet}. In the noncommutative case, in the SI period, a behavior similar to KED occurs, in which $H_{max}$ increases considerably, to later have a decrease and join the commutative behavior, see Fig. \ref{fig:Hdp}.
			
			In summary, we can assert that, within the scope of the present investigation, the early universe predicted by the noncommutative scheme considered here does not deviates in an essential way from the one predicted by the standard (effective) LQC. That is, {\it i}) the singularity is resolved via a single minimum-volume quantum bounce; {\it ii}) solutions exist that feature a sufficiently long inflationary period; {\it iii}) the different evolution stages pertaining to the pre-inflationary dynamics are in essential agreement with their standard LQC counterparts (the only meaningful effect being that of extending the time duration of the different stages). This opens the possibility of seriously considering a study related to the more fundamental level of primordial fluctuations. It could very well be the case that noncommutativity enters in a non-trivial way at this level, prompting meaningful deviations which could help temper tensions currently featured (at this level) within the standard LQC paradigm. As pointed out in Section \ref{section1}, an additional motivation (besides the more phenomenological-related ones) for considering the inclusion of a momentum-sector $\theta-$deformation at the effective scheme of LQC is that the noncommutativity among electric fluxes (momenta fields) is completely lost due to the minisuperspace approximation underlying LQC. The noncommutative scheme considered here therefore restores (some) of the non-canonical nature of the fundamental field variables of standard LQG \cite{acz}.

			\section*{Acknowledgments}
			
			This work was partially supported by the following grants: J.S.was partially supported by PROMEP UGTO-CA-3 and SNI-CONAHCyT. S. P. P. was partially supported by SNI-CONAHCyT and Secretaria de Investigación y Posgrado del Instituto Politécnico Nacional, grant SIP20231773. A. E. G. was partially supported by SNI-CONAHCyT and Secretaria de Investigación y Posgrado del Instituto Politécnico Nacional, grant SIP20231739. L. R. D. B. was partially supported by SNI-CONAHCyT and Secretaria de Investigación y Posgrado del Instituto Politécnico Nacional, grant SIP20230114. 
			

		\end{document}